\documentclass[a4paper,11pt]{article}

\usepackage{jcappub} 

\usepackage[T1]{fontenc} 

\usepackage{graphicx}
\usepackage{dcolumn}
\usepackage{amsmath}
\usepackage{amsfonts}
\usepackage{amssymb}
\usepackage{color}
\usepackage{float}
\usepackage{subfigure}
\usepackage{tabularx}
\usepackage{soul}
\usepackage{ulem}

\def\IJMPD{{Int. J. Mod. Phys. D} }

\def\JHEP{{JHEP} }

\def\MNRAS{{Mon. Not. R. Ast. Soc.} }

\newcolumntype{L}[1]{>{\raggedright\arraybackslash}p{#1}}
\newcolumntype{C}[1]{>{\centering\arraybackslash}p{#1}}
\newcolumntype{R}[1]{>{\raggedleft\arraybackslash}p{#1}}

\definecolor{v}{rgb}{0.6, 0.2, 0.8} 

\providecommand{\U}[1]{\protect\rule{.1in}{.1in}}

\title{Barrow Entropy Cosmology: an observational approach with a hint of stability analysis}

\author{Genly Leon$^1$}

\author{Juan Maga\~na$^2$}

\author{A. Hern\'andez-Almada$^3$}

\author{Miguel A. Garc\'ia-Aspeitia$^{4,5}$}

\author{Tom\'as Verdugo$^{6}$}

\author{V. Motta$^7$}

\affiliation{$^1$ Departamento  de  Matem\'aticas,  Universidad Cat\'olica del Norte, Avda.   Angamos  0610,  Casilla  1280  Antofagasta,  Chile}

\affiliation{$^2$Instituto de Astrof\'isica \& Centro de Astro-Ingenier\'ia, Pontificia Universidad Cat\'olica de Chile, \\Av. Vicu\~na Mackenna, 4860, Santiago, Chile}

\affiliation{$^3$Facultad de Ingenier\'ia, Universidad Aut\'onoma de Quer\'etaro, Centro Universitario Cerro de las Campanas, 76010, Santiago de Quer\'etaro, M\'exico.}

\affiliation{$^4$Unidad Acad\'emica de F\'isica, Universidad Aut\'onoma de Zacatecas, Calzada Solidaridad esquina con Paseo a la Bufa S/N C.P. 98060, Zacatecas, M\'exico.}

\affiliation{$^5$Consejo Nacional de Ciencia y Tecnolog\'ia, \\ Av. Insurgentes Sur 1582. Colonia Cr\'edito Constructor, Del. Benito Ju\'arez C.P. 03940, Ciudad de M\'exico, M\'exico.}

\affiliation{$^6$Instituto de Astronom\'ia, Observatorio Astron\'omico Nacional, Universidad Nacional Aut\'onoma de M\'exico, Apartado postal 106, C.P. 22800,  Ensenada, B.C., M\'exico}

\affiliation{$^7$Instituto de F\'isica y Astronom\'ia, Facultad de Ciencias, Universidad de Valpara\'iso, Avda. Gran Breta\~na 1111, Valpara\'iso, Chile.}

\emailAdd{genly.leon@ucn.cl}
\emailAdd{aldebaran.99@gmail.com}
\emailAdd{ahalmada@uaq.mx}
\emailAdd{aspeitia@fisica.uaz.edu.mx}
\emailAdd{tomasv@astro.unam.mx}
\emailAdd{veronica.motta@uv.cl}

\abstract{
In this work, we use an observational approach and dynamical system analysis to study the cosmological model recently proposed by Saridakis (2020), which is based on the modification of the entropy-area black hole relation proposed by Barrow (2020). The Friedmann equations governing the dynamics of the Universe under this entropy modification can be calculated through the gravity-thermodynamics conjecture. We investigate two models, one considering only a matter component and the other including matter and radiation, which have new terms compared to the standard model sourcing the late cosmic acceleration. A Bayesian analysis is performed in which we use five cosmological observations (observational Hubble data, type Ia supernovae, HII galaxies, strong lensing systems, and baryon acoustic oscillations) to constrain the free parameters of both models. From a joint analysis, we obtain constraints that are consistent with the standard cosmological paradigm within $2\sigma$ confidence level. In addition, a complementary dynamical system analysis using local and global variables is developed which allows obtaining a qualitative description of the cosmology. As expected, we found that the dynamical equations have a de Sitter solution at late times.}

\begin{document}
\date{\today}
\maketitle
\flushbottom
\section{Introduction}
In the last decades, one of the puzzles in Cosmology is the source of the accelerated expansion of the Universe at late times. The first observational evidence of such expansion comes from the high redshift type Ia supernovae (SNIa) \cite{Riess:1998}, confirmed by the acoustic peaks of the Cosmic Microwave Background Radiation (CMB) \cite{Aghanim:2018}, and recently tested with large scale structure measurements \cite{Nadathur:2020kvq}. The evidence point out to the existence of a dark entity whose gravitational influence should be repulsive, being known in the community with the name of dark energy (DE). The first approach, and the most successful DE candidate to explain the Universe acceleration is the cosmological constant (CC) \cite{Carroll:2000}, whose introduction in the dynamical equations is simple and in agreement with the different cosmological data and can be deduced mathematically from the Lovelock theorem \cite{Lovelock}. Despite that CC is a successful model, understanding its nature eludes us. Our best theoretical models break down under the assumption that it comes from  quantum vacuum fluctuations, obtaining results that are in total disagreement with our observations (see for example \cite{Weinberg,Zeldovich:1968ehl}). In this vein that we have been forced to propose other alternatives to explain the Universe acceleration, which is the reason behind the expression 'dark energy'. Another path to address the cosmic acceleration problem is modifying the General Theory of Relativity (GR) by assuming the DE is caused by either some geometrical effect (see the following compilation of models \cite{Garcia-Aspeitia:2016kak,Garcia-Aspeitia:2018fvw,Garcia-Aspeitia:2019yni,Garcia-Aspeitia:2019yod, review:universe}) or a fluid with strange characteristics, such as the Equation of State (EoS) taking the form $\omega<-1/3$, which is nonstandard for baryonic matter or even for dark matter (DM) (see also the models \cite{Hernandez-Almada:2018osh,Hernandez-Almada:2020ulm,Hernandez-Almada:2020uyr}).

An interesting alternative to tackle the problem of the cosmic acceleration, comes from the seminal ideas on black hole physics by Hawking and Bekenstein \cite{Bekenstein}, and hereafter applied to the cosmological context (see for instance \cite{Jacobson:1995ab,Cai:2005ra}). The formalism, known as  \textit{gravity-thermodynamics} \cite{Cai:2005ra}, consist on deriving the Einstein equations from a thermodynamic approach by using the proportionality of entropy
and horizon area, and the assumption of local equilibrium conditions.
In a recent study inspired by the geometrical structure of the COVID-19 virus, \cite{Barrow:2020tzx} propose that the expected black hole surface can be increased at the quantum gravitational level if it has such an intricate structure that could cut-off down to small scale (for instance the Planck length). In this context, Barrow constructs a fractal horizon surface by increasing the black
hole area ($A$), hence modifying its entropy as $S_{B}\sim A^{f(\Delta)}$, being $\Delta$ a constant exponent. By using the gravity-thermodynamics approach, \cite{Saridakis:2020lrg} calculated the equations governing the cosmological evolution assuming the Barrow entropy $S_B$. The modified Friedmann equation contains extra terms encoded as an effective dark energy, which drives the late cosmic acceleration. An interesting feature of this scenario is that, although the effective dark energy can behave as quintessence-like or phantom-like at different epochs, the Universe dynamics converges to de Sitter solution at larger times. By applying the Holographic principle, \cite{Saridakis:2020zol} calculated the equation governing the cosmological dynamics under the assumption that holographic dark energy obey the Barrow entropy, showing that it can source the cosmic accelerated expansion. Ref. \cite{Anagnostopoulos:2020ctz} provide observational constraints on the Barrow holographic dark energy using SNIa and measurements of the Hubble data. Later on, \cite{Mamon:2020spa} investigate the evolution of an interacting holographic dark energy model component under the Barrow's modified entropy. Recently, \cite{Adhikary:2021xym} showed that a non-flat Barrow interacting holographic dark energy can reproduce the thermal history of the Universe. In addition, the authors claim that an open Universe favors an phantom regime for the effective dark energy equation of state.

Our aim is to revisit the framework of the Barrow cosmological model proposed by \cite{Saridakis:2020lrg} to investigate the viability of such scenario to explain the late cosmic acceleration without a dark energy fluid. We constrain this model with several cosmological data at different scales: observational Hubble data, type Ia supernovae, HII galaxies, strong lensing systems, and baryonic acoustic oscillations. We also perform a dynamical analysis of the system equations to identify and classify the critical points and their stability, considering that the Universe is composed just by dust matter and filled with matter and radiation.  

The paper is organized as follow: Sec. \ref{sec:cosmo} presents the theoretical framework for the Barrow background cosmology. In Section \ref{sec:data} we  perform a Bayesian analysis to constrain the free parameters of the model using observational Hubble data, type Ia supernovae, HII galaxies, strong lensing systems and baryon acoustic oscillations. In Sec. \ref{sec:SA} we perform the  dynamical analysis and stability of the system around the critical points. Finally,  we discuss and present a summary of our results in Sec.\ref{sec:Con}. In what follows we use units in which $\hbar=k_B=c=1$.

\section{Cosmology with Barrow Entropy} \label{sec:cosmo}

The equations that govern the dynamics of the Universe can be obtained from the \textit{gravity-thermodynamics} conjecture, particularly, the Friedmann equations are retrieved by applying the first law of thermodynamics ($-dE=T\,dS$) to the apparent horizon of a Friedmann-Lemaitre-Robertson-Walker (FLRW) universe \cite{Cai:2005ra}.
Analogously to black holes whose temperature and entropy are related to its horizon area $A$, one can assume that this principle holds for the apparent cosmological horizon, $r_{A}$, i.e. it has an associated temperature $T$ and entropy $S$ in the form $T=1/2\pi r_{A}$ and $S=A/4G$, where $G$ is the Newton constant.
The heat flow (energy flux) $\delta Q$ through the horizon is given by

\begin{equation}
\delta Q=-dE= A\left(\rho_{f}+p_{f}\right) \, H\,r_{A} dt,
\end{equation}
being $\rho_f$ and $p_f$ the energy density and pressure of the fluid respectively, and the Hubble parameter at scale factor $a$ is defined as $H=\dot{a}/a$ . 
The radius of the apparent cosmological horizon $r_A$ is defined as

\begin{equation}
r_{A}=\left(H^{2}+ \frac{k}{a^2}\right)^{-1/2},
\label{eq:rA}
\end{equation}
where $k$ is the spatial Universe curvature. 

Recently, \cite{Barrow:2020tzx} propose an interesting modification to the entropy-area black hole relation by considering that the black hole horizon surface has a fractal structure. If the surface varies proportional to the radius as $\propto r^{2+\Delta}$, it modifies its entropy as

\begin{equation}
S_{B}=\left(\frac{A}{A_{0}}\right)^{1+\frac{\Delta}{2}}
\label{eq:SB}
\end{equation}
where $A$ is the standard horizon area, $A_{0}$ is the Planck area, and $\Delta$ is an exponent in the range $0<\Delta<1$. This exponent quantifies quantum deformations, when $\Delta=1$ the deformation is maximum and the Bekenstein entropy is recovered when $\Delta=0$.

Following the gravity-thermodynamics approach and assuming the Barrow entropy $S_{B}$ (Eq. \ref{eq:SB}), it is possible to obtain the Friedmann equations governing the cosmic dynamics (see further details in \cite{Saridakis:2020cqq,Saridakis:2020lrg}).
We investigate the background Cosmology in two cases: when the Universe is filled just by matter (Model I) and by matter plus radiation (Model II). In the following, we introduce the Friedmann equation in both scenarios. 

\subsection{Model I: Matter and an effective dark energy}
For a flat Universe ($k=0$) filled by matter, the Friedmann and Raychaudhuri equations \cite{Saridakis:2020lrg} are 
\begin{eqnarray}
\label{ec:friedman}
&&H^2=\frac{8\pi G}{3}\left(\rho_m+\rho_{DE}\right),\\
&&\dot{H}=-4\pi G \left(\rho_m+p_m+\rho_{DE}+p_{DE}\right),
\label{ec:raychaudhuri}
\end{eqnarray}
where $\rho_m$ denotes the energy density of matter (baryons plus dark matter) and we assume the equation of state $p_m=0$ corresponding to dust (pressureless) matter. The energy density and pressure of this effective dark energy are written in the form
 \begin{eqnarray}
&&
\rho_{DE}=\frac{3}{8\pi G} 
\left\{ \frac{\Lambda}{3}+H^2\left[1-\frac{ \beta (\Delta+2)}{2-\Delta} 
H^{-\Delta}
\right]
\right\},
\label{ec:rhoDE}
\end{eqnarray}
\begin{eqnarray}
&& 
p_{DE}= -\frac{1}{8\pi G}\Big\{
\Lambda+2\dot{H}\Big[1-\beta\Big(1+\frac{\Delta}{2}\Big) H^{-\Delta}
\Big] +3H^2\Big[1- \frac{\beta(2+\Delta)}{2-\Delta}H^{-\Delta}
\Big]\Big\}, 
\label{ec:PDE}
\end{eqnarray}
where  $ \Lambda  \equiv 4{C}G(4\pi)^{\Delta/2}$, $C$ is an appropriate integration constant and 
$\beta\equiv 4(4\pi)^{\Delta/2}G/A_{0}^{1+\Delta/2}$.
Firstly, we keep $\beta$ as a free parameter. Next, by setting $A_{0}=4G$ we have $\beta\equiv  ( \pi/G)^{\Delta/2}$. 

Although Eqs. \eqref{ec:friedman}-\eqref{ec:raychaudhuri}  does not has a dark energy component, we have dubbed \textit{effective dark energy} the extra-terms introduced by the Barrow entropy. 

The equation of state (EoS) for the effective dark energy reads
\begin{eqnarray}
w_{DE}\equiv\frac{p_{DE}}{\rho_{DE}}=-1-
\frac{     
  2\dot{H}\left[1-\beta\left(1+\frac{\Delta}{2}\right) H^{-\Delta}
\right]
 }{\Lambda+3H^2\left[1-\frac{\beta(2+\Delta)}{2-\Delta}H^{-\Delta}
\right]}
\label{FRWwDE}.
\end{eqnarray}
Combining \eqref{ec:friedman} and 
\eqref{ec:rhoDE}, 
the dimensionless Friedmann equation takes the form
\begin{equation}
\label{normalized:Friedman}
 \frac{8 \pi  G \rho_{m}}{3 H^2}+\frac{\Lambda }{3 H^2}=\frac{\beta  (\Delta +2)
   H^{-\Delta }}{2- \Delta}.   
\end{equation}
The standard model with cold dark matter and cosmological constant ($\Lambda$CDM) is recovered with $\Delta=0, \beta=1$, which implies  
\begin{equation}
    \frac{8 \pi  G \rho_{m}}{3 H^2}+\frac{\Lambda }{3 H^2}=1.
\end{equation}
Now, we define 
\begin{align}
& \Omega_{m} \equiv   \frac{8 \pi  G \rho_{m}}{3 H^2}, \quad \Omega_\Lambda\equiv  \frac{\Lambda }{3 H^2}, \nonumber\\
&  \Omega_{DE} \equiv  \frac{8 \pi  G \rho_{DE}}{3 H^2} = 1 + \Omega_\Lambda - \frac{\beta  (\Delta +2)
   H^{-\Delta }}{2- \Delta}. \label{New_Vars_1}
\end{align}
Then, from Eq. \eqref{normalized:Friedman} it follows
\begin{equation}
   \Omega_{m}+ \Omega_{DE}  =1+   \underbrace{ \Omega_m + \Omega_\Lambda - \frac{\beta  (\Delta +2)
   H^{-\Delta }}{2- \Delta}}_{=0}  =1. 
\end{equation}
By considering that the matter component evolves in the traditional way $\rho_m=\rho_{m0}(z+1)^3$, the dimensionless Friedmann equation can alternatively be written as
\begin{align}
 & E(z)\equiv\frac{H}{H_0}=\Big\lbrace\bar{\beta}\frac{2-\Delta}{2+\Delta}[\Omega_{m0}(z+1)^3+\Omega_{\Lambda0}]\Big\rbrace^{1/(2-\Delta)},
 \label{eq:Ez}
\end{align}
where $a=(z+1)^{-1}$, $z$ is the redshift, the subscripts $0$ denote quantities at $a=1 (z=0)$, and we have defined the dimensionless  parameter
\begin{eqnarray}
    &&\bar{\beta}\equiv\frac{H_0^{\Delta}}{\beta}=\frac{H_0^{\Delta} A_{0}^{1+\Delta/2}}{4(4\pi)^{\Delta/2}G} \underbrace{= {H_0^{\Delta} (G/\pi)^{\Delta/2}}}_{\text{setting} \; A_{0}=4G},
\label{eq:var_dimless}
\end{eqnarray}
\noindent
where we have set the Planck area $A_{0}=4G$. Notice that when $\Delta=0$, and $\bar{\beta}=1$, the $\Lambda$CDM model is recovered.

Furthermore, the flatness constraint $E(0)=1$, gives the equation
\begin{equation}
\label{eq2.15}
    \Omega_{\Lambda0}=\left(\frac{2+\Delta}{2-\Delta}\right)\frac{1}{\bar{\beta}}-\Omega_{m0}.
\end{equation}
In addition, the deceleration parameter is defined as
\begin{equation}
\label{eq:qz}
q=-1 -   \frac{\dot{H}}{H^2}.
\end{equation}
By substituting Eq. \eqref{eq:Ez} and replacing Eqs. \eqref{ec:rhoDE}, \eqref{ec:PDE} into \eqref{ec:raychaudhuri}, the $q(z)$ results
\begin{eqnarray}
    &&q(z)=-1 +\frac{3\Omega_{m0}(z+1)^3}{E(z)} \left[\frac{\bar{\beta}(2-\Delta)^{\Delta-1}}{(2+\Delta)}\right]^{1/(2-\Delta)}[\Omega_{m0}(z+1)^3\nonumber+ \Omega_{\Lambda0}]^{(\Delta-1)/(2-\Delta)},\\
\end{eqnarray}
and the DE EoS in terms of Eqs. \eqref{eq:Ez}-\eqref{eq:var_dimless} is written as

\begin{eqnarray}
&&w_{DE}=-1+\frac{2(z+1)E(z)E'(z)[1-(1+\Delta/2)(E\bar{\beta})^{-\Delta}]}{3\Omega_{\Lambda}+3E(z)^2\left[1-\frac{(2+\Delta)}{2-\Delta}(E\bar{\beta})^{-\Delta}\right]}
\label{eq:wde}
\end{eqnarray}
where
\begin{eqnarray}
E'(z)=3\Omega_{m0}\frac{\bar{\beta}}{2+\Delta}E(z)^{(\Delta-1)}(z+1)^2.
\end{eqnarray}
It is worthy to note that for $\bar{\beta}=1$, $\Delta=0$ and $z\to1$ the EoS for the cosmological constant is recovered, i.e. $w_{DE}=-1$.

\subsection{Model II: Matter, Radiation and
Effective Dark Energy} 

By considering two fluids, matter and radiation, for a flat Universe, the Friedmann and Raychaudhuri equations in Barrow cosmology result as follows
\begin{eqnarray}
\label{ec:friedman2}
&&H^2=\frac{8\pi G}{3}\left(\rho_m+\rho_r+\rho_{DE}\right),\\
&&\dot{H}=-4\pi G \left(\rho_m+p_m+\rho_r+p_r+\rho_{DE}+p_{DE}\right),
\label{ec:raychaudhuri2}
\end{eqnarray}
where $\rho_r$ indicates the radiation energy density and the radiation pressure is $p_r=\rho_r/3$. We define 
\begin{align}
\label{varOmega_r}
& \Omega_{r} \equiv   \frac{8 \pi  G \rho_{r}}{3 H^2}.
\end{align}

Using the previous variable definitions \eqref{New_Vars_1},  the dimensionless Friedmann equation becomes \begin{equation}
    \label{normalized:Friedman2_2}
 \Omega_{\Lambda}+\Omega_m + \Omega_r=\frac{\beta  (\Delta +2)
   H^{-\Delta }}{2- \Delta}.  
\end{equation} 
Finally, from Eq. \eqref{normalized:Friedman2_2} it follows 
\begin{equation}
   \Omega_{m}+ \Omega_{r}+ \Omega_{DE}  =1+   \underbrace{\Omega_m + \Omega_{r} + \Omega_\Lambda - \frac{\beta  (\Delta +2)
   H^{-\Delta }}{2- \Delta}}_{=0}  =1. 
\end{equation}

The dimensionless Friedmann equation can be written as
\begin{eqnarray}
 E(z)\equiv\frac{H}{H_0}=\Big\lbrace\bar{\beta}\frac{2-\Delta}{2+\Delta}[\Omega_{m0}(z+1)^3+\Omega_{r0}(z+1)^4+\Omega_{\Lambda0}]\Big\rbrace^{1/(2-\Delta)},
 \label{eq:Friedmann_matrad}
\end{eqnarray}
where the radiation component evolves in the traditional way $\rho_r=\rho_{r0}(z+1)^4$, and $\Omega_{r0}$ is obtained from \eqref{varOmega_r} evaluated at $a=1$ ($z=0$). This can be calculated as $\Omega_{r0}=2.469 \times 10^{-5}h^{-2} (1+0.2271N_{eff})$, with $N_{eff}=3.04$ as the number of relativistic species \citep{Komatsu:2011} and $h=H_{0}/ 100 \mathrm{km s}^{-1}\mathrm{Mpc}^{-1}$ as the current Hubble dimensionless parameter.

The flatness constriction $E(0)=1$, gives the equation
\begin{equation}
\label{flatness2}
    \Omega_{\Lambda0}=\left(\frac{2+\Delta}{2-\Delta}\right)\frac{1}{\bar{\beta}}-\Omega_{m0}-\Omega_{r0}.
\end{equation}
In addition, the deceleration parameter reads
\begin{eqnarray}
    &&q(z)=-1 + \left[\frac{\bar{\beta}(2-\Delta)^{\Delta-1}}{(2+\Delta)}\right]^{1/(2-\Delta)}\times\nonumber\\&&
    \frac{3\Omega_{m0}(z+1)^3+4\Omega_{r0}(z+1)^4}{E(z)}[\Omega_{m0}(z+1)^3+\Omega_{r0}(z+1)^4+\Omega_{\Lambda0}]^{(\Delta-1)/(2-\Delta)}.
\end{eqnarray}
The DE EoS can be calculated by substituting Eq. \eqref{eq:Friedmann_matrad} into Eq. \eqref{eq:wde} and $E'(z)$ as 
\begin{eqnarray}
&&E'(z)=\frac{\bar{\beta}}{2+\Delta}E(z)^{(\Delta-1)}[3\Omega_{m0}(z+1)^2+4\Omega_{r0}(z+1)^3].
\end{eqnarray}
Notice that for $\bar{\beta}=1$, $\Delta=0$ and $z\to1$ the EoS for the cosmological constant is recovered.

\section{Observational constraints} \label{sec:data}
For both models under Barrow cosmology (Model I: universe filled by matter and Model II: universe filled by matter plus radiation), the free parameters of the model are: $h$, $\Omega_{m0}$, $\bar{\beta}$, and $\Delta$.
To constrain these parameters we employ observational Hubble data (OHD) \cite{Magana:2017nfs}, Type Ia supernovaes (SNIa) \cite{Scolnic:2017caz}, HII galaxies (HIIG) \cite{Cao:2020jgu}, strong lensing systems (SLS) \cite{Amante:2019xao}, and baryon acoustic oscillations (BAO) \cite{Nunes:2020hzy}. In
the following we briefly describe these samples.

\subsection{Observational Hubble Data}

A cosmological-independent measurement of the Hubble parameter is acquired through the differential age (DA) technique \citep{Moresco:2016mzx} in cosmic chronometers (i.e passive elliptic galaxies).
In this paper, we consider the OHD compilation provided by \cite{Magana:2017nfs} containing $31$ data points in the range $0.07<z<1.965$. Hence, the chi square function for OHD can be constructed through the following equation
\begin{equation}
    \chi^2_{{\rm OHD}}=\sum_i^{31}\left(\frac{H_{th}({\bf\Theta},z_i)-H_{obs}(z_i)}{\sigma^i_{obs}}\right)^2,
\end{equation}
where $H_{th}({\bf\Theta},z_i)$ and $H_{obs}(z_i)$, are the theoretical and observational Hubble parameters respectively at the redshift $z_i$, and $\sigma^i_{obs}$ is the observational error. Notice that ${\bf\Theta}$ is a vector related to the number of free parameters of the studied cosmological model.

\subsection{Pantheon SNIa sample}

The Pantheon sample \cite{Scolnic:2018} contains $1048$ SNIa data points in the redshift range $0.001<z<2.3$.
The observational distance modulus $\mu_{\mathrm{PAN}}$ for Pantheon SNIa can be measured as

\begin{equation}
  \mu_{\mathrm{PAN}} = m_{b}^{\star} - M_{B} + \alpha \times X_{1} - \beta \times C + \Delta_M+ \Delta_B, 
  \label{eq:mu}
\end{equation}
\noindent
where $m_{b}^{\star}$ corresponds to the observed peak magnitude, $M_B$ is the B-band absolute magnitude, $\alpha$, and $\beta$ coefficients are nuisance parameters;
$X_{1}$ and $C$ are variables describing the time stretching of the light-curve and the Supernova color at maximum brightness, respectively. 
$\Delta_M$ is a distance correction based on the host-galaxy mass of the SNIa and $\Delta_B$ is a distance correction based on predicted biases from simulations.  

The theoretical counterpart of the distance modulus for any cosmological model is given by $\mu_{th}({\bf\Theta},z) = 5 \log_{10} \left( d_L({\bf\Theta},z) / 10 \mathrm{pc} \right)$, where $d_{L}$ is the luminosity distance given by
 \begin{equation}
d_{L}({\bf\Theta},z)=\frac{c}{H_{0}}(1+z) \int^{z}_{0}\frac{\mathrm{dz}^{\prime}}{E(z^{\prime})},
\label{eq:dl}
\end{equation}
where $c$ is the light speed velocity.
Since that \cite{Scolnic:2018} provide
$\tilde{\mu}_{\mathrm{PAN}}=\mu_{\mathrm{PAN}}+M_{B}$, we can marginalize over the $M_{B}$ parameter to compare the data with the underlying cosmology. Thus, the marginalized figure-of-merit for the Pantheon sample is given by
\begin{equation}
\chi_{Pan_{Mmarg}}^{2}=a +\log \left( \frac{e}{2\pi} \right)-\frac{b^{2}}{e}, \label{fPan}
\end{equation}
\noindent
where $a=\Delta\boldsymbol{\tilde{\mu}}^{T}\cdot\mathbf{C_{P}^{-1}}\cdot\Delta\boldsymbol{\tilde{\mu}},\, b=\Delta\boldsymbol{\tilde{\mu}}^{T}\cdot\mathbf{C_{P}^{-1}}\cdot\Delta\mathbf{1}$,\, $e=\Delta\mathbf{1}^{T}\cdot\mathbf{C_{P}^{-1}}\cdot\Delta\mathbf{1}$, and $\Delta\boldsymbol{\tilde{\mu}}$ is the vector of residuals between the model distance modulus and the observed $\tilde{\mu}_{\mathrm{PAN}}$. The covariance matrix $\mathbf{C_{P}}$ is constructed by adding the systematic and statistical matrices of $\tilde{\mu}_{\mathrm{PAN}}$. We refer the interested reader to \cite{Scolnic:2018} for a detailed description of how these matrices are constructed.

\subsection{HII Galaxies}

HIIG are galaxies with large HII regions, product of young and hot stars (O and/or B type stars) ionizing the medium. For these galaxies there is a correlation between the measured luminosity, $L$,  and the inferred velocity dispersion, $\sigma$, of the ionized gas. Several authors have shown that the correlation $L-\sigma$ could be used as a cosmological tracer \citep[][and references therein]{Chavez2012,Chavez2014,Terlevich2015,Chavez2016,GonzalezMoran2019}. A HIIG data sample was compiled by \cite{GonzalezMoran2019} containing $107$ low redshift ($0.0088$ $\leq$ $z$ $\leq$ $0.16417$)  galaxies, and $46$ high redshift ($0.636427$ $\leq$ $z$ $\leq$ $2.42935$) galaxies.  \cite{Cao:2020jgu} used such HIIG sample to constrain the cosmological parameters for six different cosmological models. 
Recently, \cite{Gonzalez-Moran:2021drc} presented a new sample which contains $181$ local and high-z HIIG data points in the redshift range $0.01<z<2.6$. In this paper, we use such HIIG sample, and follow their methodology  \cite[see][]{Gonzalez-Moran:2021drc}.

The correlation between $L$ and $\sigma$ can be written as
\begin{equation}
    \log L=\beta_{II}\log\sigma+\gamma,
\end{equation}
where $\gamma$ and $\beta_{II}$ are the intercept and slope functions, respectively. Following \cite{Cao:2020jgu,GonzalezMoran2019, Gonzalez-Moran:2021drc}, we set $\beta_{II}$ = 5.022, and $\gamma$ = 33.268. Therefore, the distance modulus takes the form
\begin{equation}
    \mu_{obs}=2.5\log L-2.5\log f-100.2,
    \label{eq:muobs_hii}
\end{equation}
where $f$ is the flux emitted by the HIIG. Moreover, the theoretical distance modulus is
\begin{equation}
    \mu_{th}({\bf\Theta},z)=5\log d_L({\bf\Theta},z)+25,
\end{equation}
being $d_L({\bf\Theta},z)$ the luminosity distance (in Mpc).

The figure-of-merit is given by the following equation
\begin{equation}
    \chi^2_{{\rm HIIG}}=\sum_i^{18}\frac{[\mu_{th}({\bf\Theta},z_i)-\mu_{obs}(z_i)]}{\epsilon_i^2},
\end{equation}
where $\epsilon_i$ is the uncertainty of the $i_{th}$ measurement and it can be calculated propagating the errors of Eq. \ref{eq:muobs_hii} \cite[see further details in][]{Gonzalez-Moran:2021drc}.

\subsection{Strong lensing systems}
Several authors have shown that strong lensing systems
can be used as cosmological tool to constrain cosmological parameters \cite{Amante:2019xao}. The method consists in comparing a theoretical distance ratio of angular diameter distances in the lens geometry with its observational counterpart. It can be obtained from  the Einstein radius of a lens (modeled with a singular isothermal sphere) given by
\begin{equation}
    \theta_E=4\pi\frac{\sigma_{l}^2D_{ls}}{c^2D_s},
\end{equation}
where $\sigma_{l}$ is the observed velocity dispersion of the lens galaxy, $D_s$ is the angular diameter distance to the source at redshift $z_{s}$, and $D_{ls}$ is the angular diameter distance from the lens (at redshift $z_{l}$) to the source. Then, the observational distance ratio of angular diameter distances is defined as
\begin{equation}
D^{obs}\equiv c^2\theta_E/4\pi\sigma^2.
\end{equation}
To measure the theoretical distance ratio $D^{th}({\bf\Theta},z_l,z_s)\equiv D_{ls}/D_s$, we calculate $D_{s}$ using the definition of angular diameter distance of a source at redshift $z$
\begin{equation}
         D_{A}({\bf\Theta},z)=\frac{c}{H_0 (1+z)}\int_{0}^{z}\frac{dz^{\prime}}{E(z^{\prime})},
\label{eq:DA}
\end{equation}
and $D_{ls}$ through the definition of the angular diameter distance between two objects at redshift $z_{1}$ and $z_{2}$ 
\begin{equation}
         D_{12}({\bf\Theta},z)=\frac{c}{H_0 (1+z_{2})}\int_{z_1}^{z_2}\frac{dz^{\prime}}{E(z^{\prime})}.
\end{equation}
The most recent compilation of Strong-Lensing Systems (SLS) given by \cite{Amante:2019xao} consists of 204 SLS spanning the redshift region $0.0625<z_l<0.958$ for the lens and $0.196<z_s<3.595$ for the source. To avoid convergence problems and discarding (unphysical) systems with $D_{ls}>D_{s}$, the authors provided a fiducial sample with an observational lens equation ($D^{obs}$) within the region $ 0.5 \leq D^{obs} \leq 1$. 

In this work, we use such a fiducial sample consisting of 143 SLS, and the chi-square function takes the form
\begin{equation}
    \chi^2_{\rm SLS}=\sum_i^{143}\frac{[D^{th}({\bf\Theta},z_l,z_s)-D^{obs}(\theta_E,\sigma^2)]^2}{(\delta D^{obs})^2},
\end{equation}
where 
\begin{equation}
    \delta D^{obs}=D^{obs}\left[\left(\frac{\delta\theta_E}{\theta_E}\right)^2+4\left(\frac{\delta\sigma}{\sigma}\right)^2\right]^{1/2},
\end{equation}
being $\delta\theta_E$ and $\delta\sigma$ the uncertainties of the Einstein radius and velocity dispersion, respectively.

\subsection{Baryon Acoustic Oscillations}

BAO are considered as  standard rulers, being primordial signatures of the interaction of baryons and photons in a hot plasma on the matter power spectrum in the pre-recombination epoch. Authors in \cite{Giostri:2012} collected 6 correlated data points measured by \cite{Percival:2010,Blake:2011,Beutler:2011hx}. 
To confront cosmological models to these data, it is useful to build the $\chi^2$-function in the form
\begin{equation}
\chi^2_{\rm BAO} =  \Delta\Vec{X}^T  \rm{Cov}^{-1} \Delta \vec{X} \,,
\end{equation}
where $\Delta \Vec{X}$ is the difference between the theoretical and observational values of $d_A(z_{drag})/D_V(z_i)$ where $z_{drag}$ is defined by the sound horizon at baryon drag epoch measured at the redshift $z_i$, and  $\rm{Cov}^{-1}$ is the inverse of covariance matrix, the dilation scale ($D_V$) is defined as \cite{Wigglez:Eisenstein2005}
\begin{equation} \label{eq:D_V}
    D_V ({\bf\Theta},z)= \left[ \frac{d_A^2({\bf\Theta},z)\,c\,z}{H_0 E(z)}\right]^{1/3}
\end{equation}
where $d_A({\bf\Theta},z)=(1+z)D_A({\bf\Theta},z)$ is the comoving angular-diameter distance and $D_A$ is the angular diameter distance at $z$ presented in Eq. \eqref{eq:DA}. Additionally, $r_{drag}$ is the sound horizon at baryon drag epoch. We use $r_{drag}=147.21 \pm 0.23$ reported in \cite{Aghanim:2018}.

\subsection{Results from Observational constraints} 
The inference of the cosmological parameters under Barrow  cosmology, for both model I (Eq. \ref{eq:Ez}) and II (Eq. \ref{eq:Friedmann_matrad}), is performed using a Bayesian  Markov Chain Monte Carlo (MCMC) approach using the emcee Python module \cite{Emcee:2013}. We set $3000$ chains with $250$ steps each one. The burn-in phase is stopped up to obtain convergence according to the auto-correlation time criteria. 
We build a Gaussian log-likelihood as the merit-of-function to minimize through the equation
$-2\log(\mathcal{L}_{\rm data})\varpropto \chi^2_{\rm data}$ 
for each dataset, and consider Gaussian priors on $h$ and $\Omega_{m0}$ centered at $0.6766\pm 0.0042$ and $0.3111\pm 0.0056$ \cite{Aghanim:2018}, respectively, and a flat prior for $\Delta$ in the ranges: $\Delta:\,[0,2]$. The parameter $\bar\beta$ is calculated using Eq. \eqref{eq:var_dimless}, where we have set $A_{0}=4G$. Additionally, a joint analysis can be constructed through the sum of their function-of-merits, i.e.,
\begin{equation}
    \chi^2_{\rm Joint}=\chi^2_{\rm SNIa}+\chi^2_{\rm BAO}+\chi^2_{\rm OHD}+\chi^2_{\rm SLS}+\chi^2_{\rm HIIG},
\end{equation}
where subscripts indicate the observational measurements under consideration.
 
\begin{figure}
    \centering
    \includegraphics[width=0.65\textwidth]{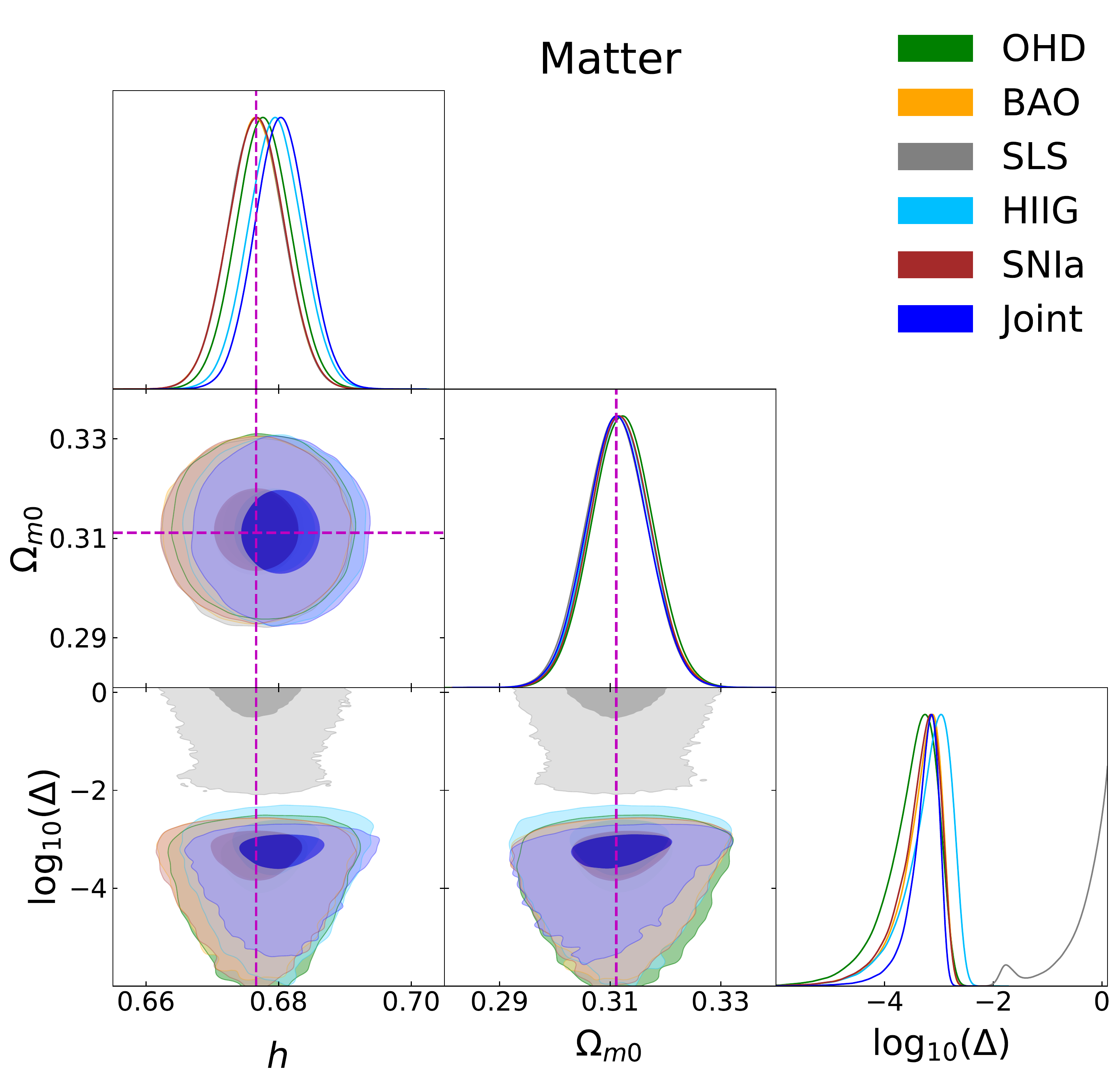}\\
    \includegraphics[width=0.65\textwidth]{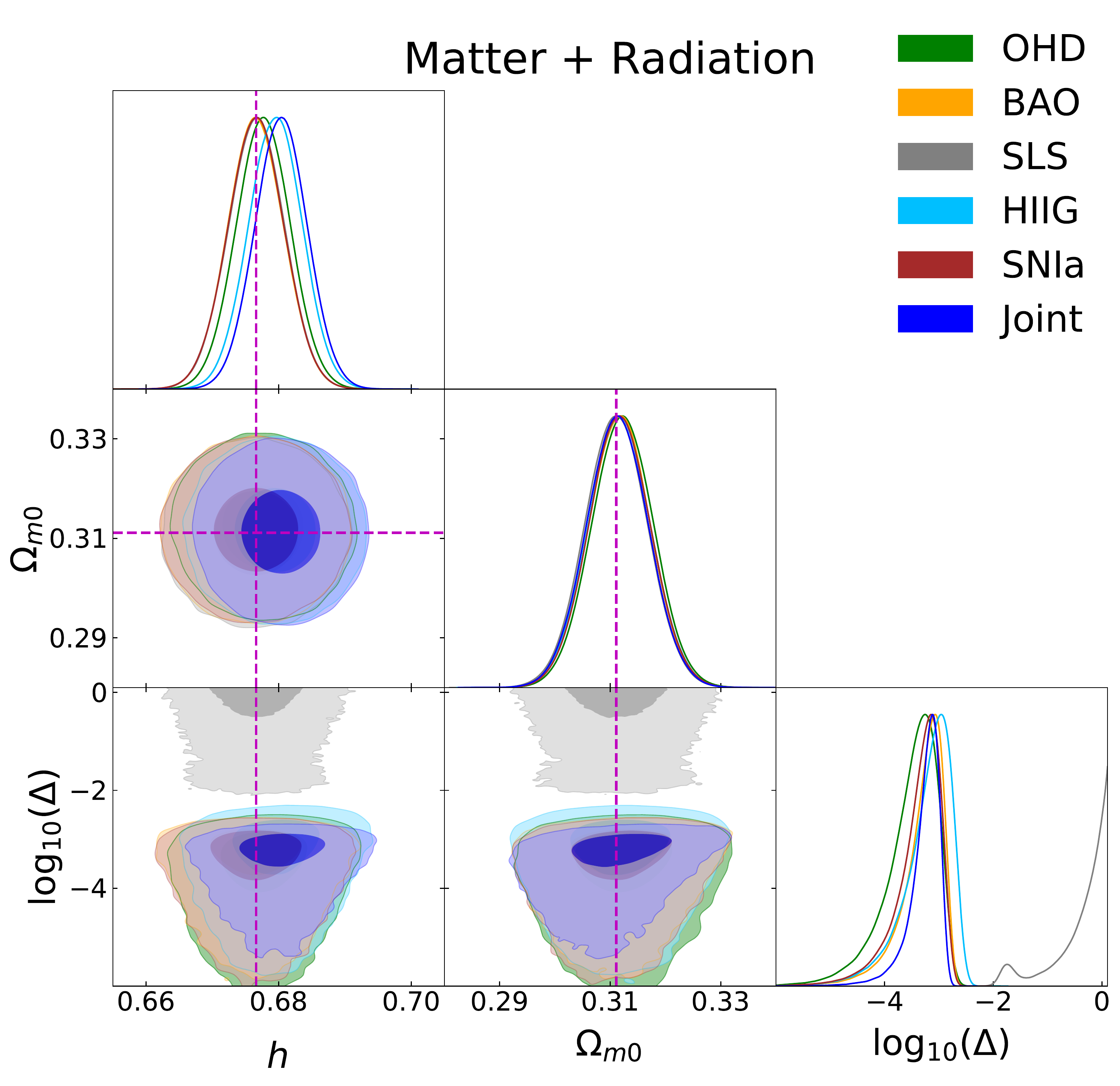}
    \caption{2D contour at 68\% and 99.7\% CL and 1D posterior distribution of the free parameters in two models: universe filled by matter (top panel) and universe filled by matter plus radiation (bottom panel). Dashed lines represent best-fit values for $\Lambda$CDM \cite{Aghanim:2018}.}
    \label{fig:contours}
\end{figure}

Figure \ref{fig:contours} shows the 2D confidence contours at $68\%$ ($1\sigma$) and $99.7\%$ ($3\sigma$) confidence level (CL) respectively, and 1D posterior distribution of the parameters in Barrow cosmology with a matter component (top panel) and matter plus radiation. In the case of the $\Delta$ parameter, although the contours for most of the samples are consistent with each other, the ones obtained using SLS data are in tension with those estimated with the other samples.  However, this is not surprising inasmuch as reported by \cite{Amante:2019xao}, the use of their fiduciary sample of 143 strong lensing systems while performs better constraining the cosmological models tested in such work (compared with other lensing samples), the parameters are not tightly constraint. Indeed, they reported that the range on the studied cosmological parameters were in agreement with those expected from other astrophysical observations, but they also discussed that the method needs improvement, in particular to take into account systematic biases (e.g. not fully confirmed lenses, multiple arcs, uncertain redshifts, complex lens substructure).

Table \ref{tab:bestfits} presents the chi-square and mean values of the parameters obtained from the different data set and their uncertainties at $1\sigma$ for both Barrow cosmologies. We obtain $\Delta=(5.912^{+3.353}_{-3.112})\times 10^{-4}$, $\bar{\beta}=0.920^{+0.042}_{-0.042}$, and  $\Delta=(6.245^{+3.377}_{-3.164})\times 10^{-4}$, $\bar{\beta}=0.915^{+0.043}_{-0.043}$ for model I and II, respectively. Both models are consistent at $2\sigma$ with the standard cosmological model, i.e. $\Delta=0$ and $\bar{\beta}=1$. Moreover, the $\Delta$ bounds suggest the entropy-area relation is consistent with the Bekenstein entropy. From now on, we focus our discussion on the model II ( matter and radiation components) since it is a more realistic model.

\begin{figure}
\centering
\includegraphics[width=0.55\textwidth]{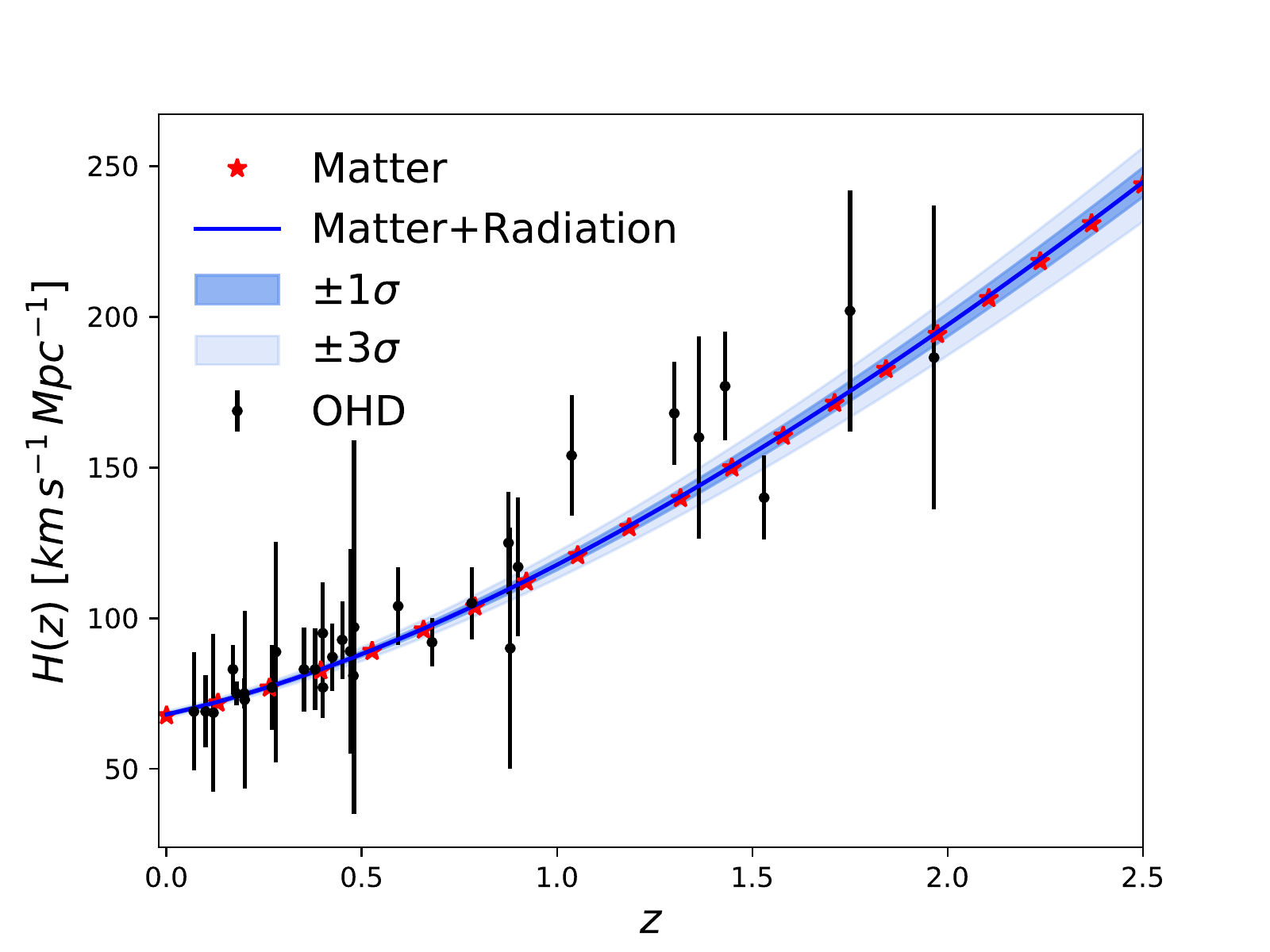}\\
\includegraphics[width=0.55\textwidth]{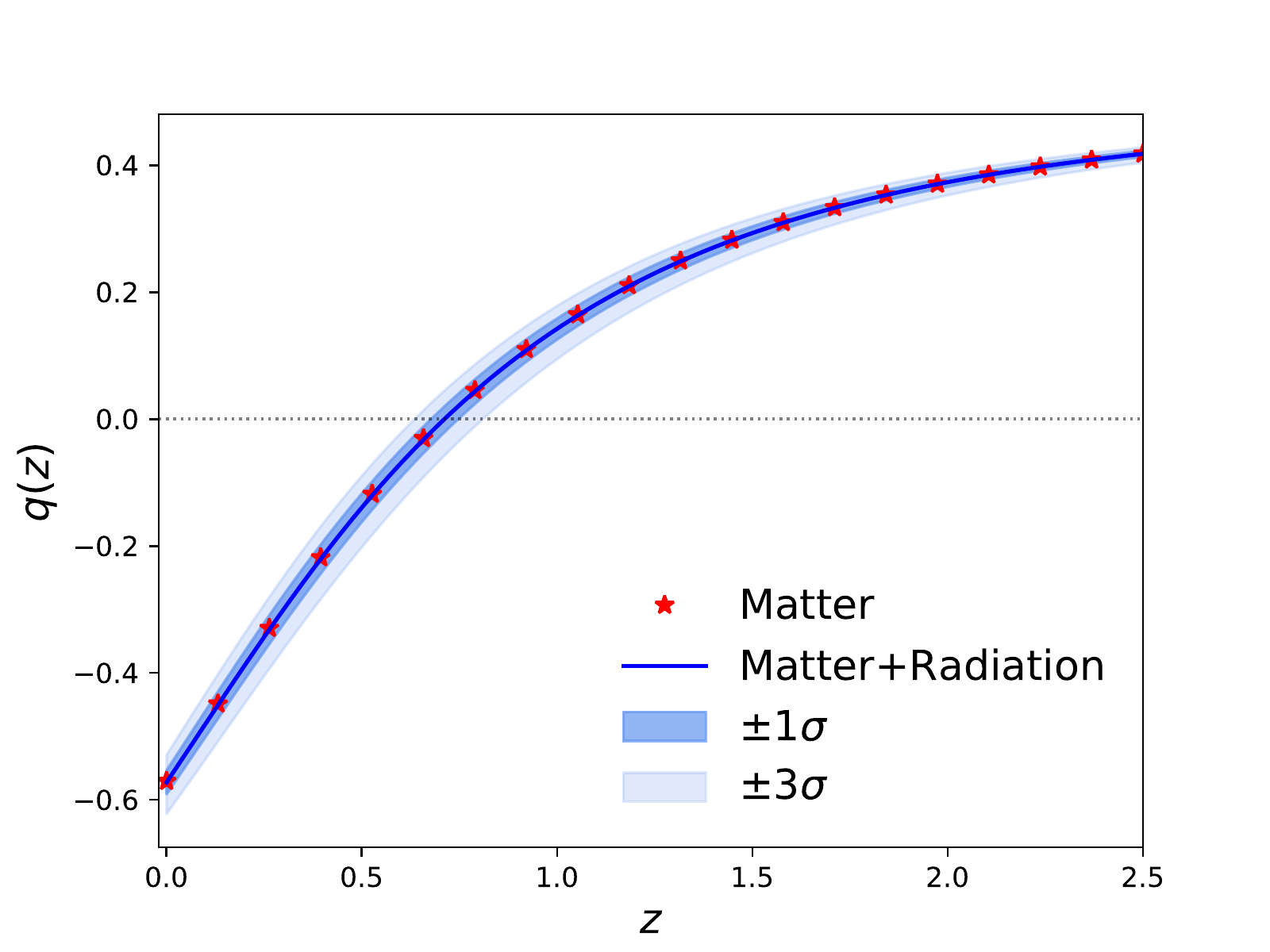}\\
\includegraphics[width=0.55\textwidth]{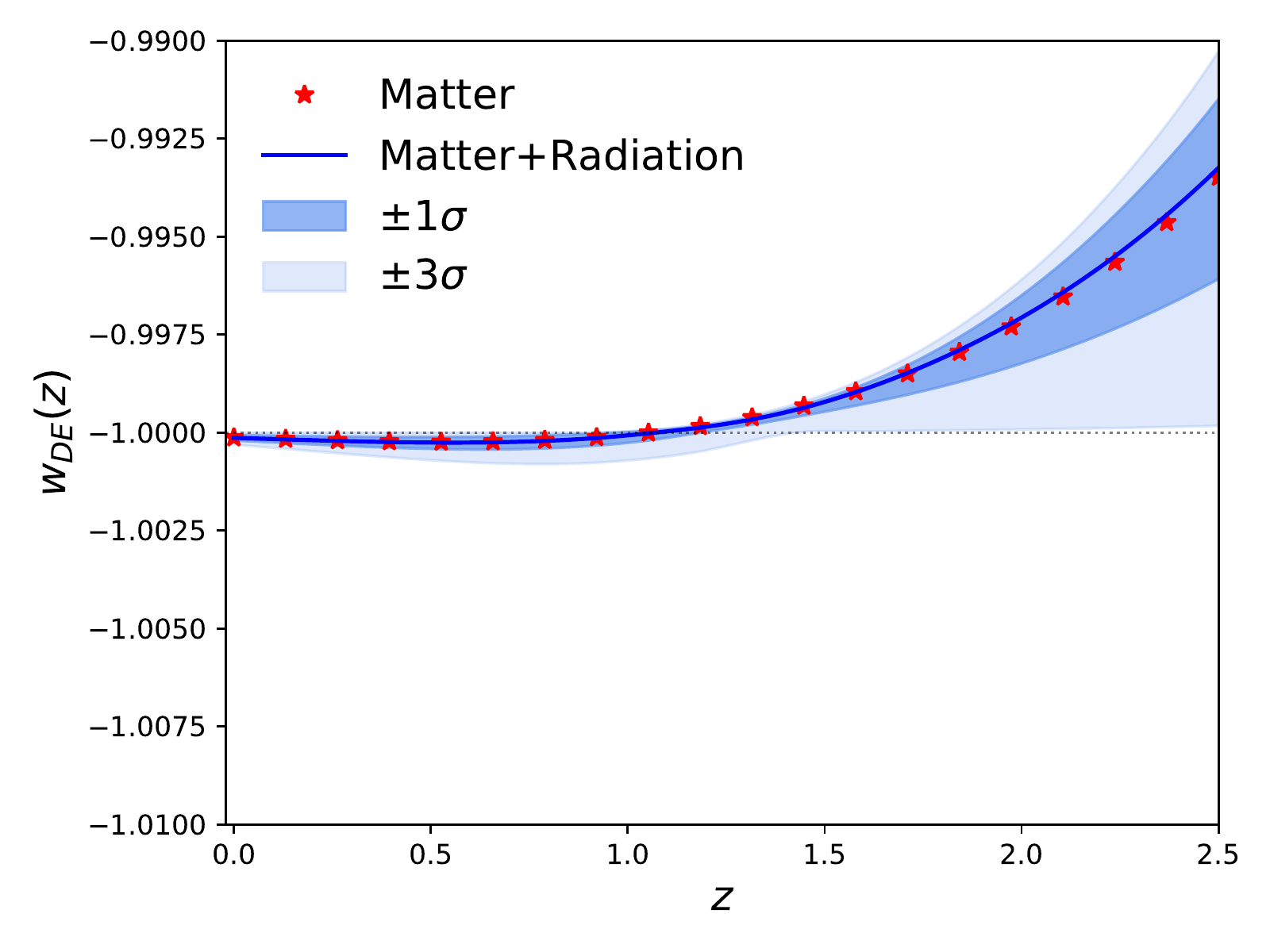}
\caption{Top panel: $H(z)$ function for the Barrow cosmological model using the mean values of the joint analysis. Middle panel: Reconstruction of the cosmological evolution of $q(z)$ function using the mean values of the joint analysis. Bottom panel : Reconstruction of the equation-of-state of the effective dark energy. In all panels the shadow regions represent the $1\sigma$, and $3\sigma$ (from darker to lighter color bands) confidence levels.}
\label{fig:Hz_and_qz}
\end{figure}

The top panel of the Figure \ref{fig:Hz_and_qz} shows that the expansion rate estimated from the mean values of the Barrow cosmology parameters are consistent with the OHD. In addition, the reconstruction of the deceleration parameter as function of redshift is shown in the middle panel of Fig. \ref{fig:Hz_and_qz}. The $q(z)$ behavior is similar to the standard one, i.e. there is a transition at $z_{t}\simeq 0.711^{+0.035}_{-0.034}$ from a decelerated stage to an accelerated stage with $q_0=-0.573^{+0.019}_{-0.019}$, suggesting a de Sitter solution. However, in the Barrow scenario the late cosmic acceleration is driven by the new terms in the dynamical equations. Finally, the bottom panel of Fig. \ref{fig:Hz_and_qz} illustrates the reconstruction of the equation of state of the effective dark energy as function of redshift. It is worth noting that it has a transition at $z_{wde}\simeq 1.070^{+0.114}_{-0.104}$ from a quintessence-like regime to a phantom-like one, yielding $w_{DE}(0)\simeq -1.000134^{+0.000069}_{-0.000068}$ at current times, which is consistent with the cosmological constant at $1\sigma$. This behavior of $w_{DE}(z)$ has been also discussed by \cite{Saridakis:2020lrg}, the effective dark energy can undergo the phantom-divide crossing but it tends asymptotically to a de Sitter solution at late times.

Furthermore, the age of the Universe can be estimated by solving the integral $t_{0}=\int_{0}^{1} {\text da}/a H(a)$. 
Considering the constraints from the joint analysis, we obtain $t_{0}\simeq 14.062^{+0.179}_{-0.170}$ ($14.045^{+0.179}_{-0.167}$)  Gyr for matter+radiation (matter) model, consistent with $2\sigma$ confidence level with the measurements of Planck  \cite{Aghanim:2018}.
Thus, the constraints obtained from several data at different scales indicate that, by modifying the entropy-area relation,  Barrow cosmology is a plausible scenario to explain the late cosmic acceleration without the need to include an exotic component. 

In the following sections, we perform a dynamical system analysis of the Barrow cosmology.

\begin{table*}
\caption{Mean values of the free parameters for the Barrow cosmology in two scenarios: universe filled by matter (Model I) and by matter plus radiation (Model II).}
\centering
\begin{tabular}{|lcccc|}
\hline
Sample     &    $\chi^2$     &  $h$ & $\Omega_m^{(0)}$ & $\Delta$  \\
\hline
\multicolumn{5}{|c|}{Model I} \\ [0.9ex]
SLS    & $217.28$  & $0.677^{+0.004}_{-0.004}$ & $0.311^{+0.006}_{-0.006}$ & $0.963^{+0.704}_{-0.700}$ \\ [0.9ex] 
BAO    & $2.97$    & $0.677^{+0.004}_{-0.004}$ & $0.312^{+0.006}_{-0.005}$ & $(5.370^{+4.713}_{-3.550})\times 10^{-4}$ \\ [0.9ex] 
DA OHD & $15.62$   & $0.678^{+0.004}_{-0.004}$ & $0.312^{+0.006}_{-0.006}$ & $(3.813^{+4.774}_{-2.752})\times 10^{-4}$ \\ [0.9ex] 
HII    & $441.64$  & $0.679^{+0.004}_{-0.004}$ & $0.312^{+0.005}_{-0.006}$ & $(7.548^{+8.094}_{-5.272})\times 10^{-4}$ \\ [0.9ex] 
SNIa   & $1036.11$ & $0.677^{+0.004}_{-0.004}$ & $0.312^{+0.005}_{-0.005}$ & $(5.056^{+4.678}_{-3.361})\times 10^{-4}$ \\ [0.9ex] 
Joint  & $1749.08$ & $0.680^{+0.004}_{-0.004}$ & $0.311^{+0.006}_{-0.005}$ & $(5.912^{+3.353}_{-3.112})\times 10^{-4}$ \\ [0.9ex] 
\hline
\multicolumn{5}{|c|}{Model II} \\ 
SLS    & $217.28$  & $0.677^{+0.004}_{-0.004}$ & $0.311^{+0.006}_{-0.006}$ & $0.963^{+0.703}_{-0.708}$ \\ [0.9ex] 
BAO    & $2.86$    & $0.677^{+0.004}_{-0.004}$ & $0.312^{+0.005}_{-0.006}$ & $(6.008^{+4.949}_{-3.880})\times 10^{-4}$ \\ [0.9ex] 
DA OHD & $15.61$   & $0.678^{+0.004}_{-0.004}$ & $0.312^{+0.006}_{-0.006}$ & $(3.809^{+4.855}_{-2.740})\times 10^{-4}$ \\ [0.9ex] 
HII    & $441.64$  & $0.680^{+0.004}_{-0.004}$ & $0.312^{+0.006}_{-0.006}$ & $(7.540^{+8.107}_{-5.245})\times 10^{-4}$ \\ [0.9ex] 
SNIa   & $1036.11$ & $0.677^{+0.004}_{-0.004}$ & $0.312^{+0.006}_{-0.005}$ & $(5.105^{+4.680}_{-3.383})\times 10^{-4}$ \\ [0.9ex] 
Joint  & $1748.89$ & $0.680^{+0.004}_{-0.004}$ & $0.311^{+0.006}_{-0.006}$ & $(6.245^{+3.377}_{-3.164})\times 10^{-4}$ \\ [0.9ex]
\hline
\end{tabular}
\label{tab:bestfits}
\end{table*}

\section{Dynamical system and stability analysis} \label{sec:SA}

The  phase-space and stability analysis is a complementary
 inspection that allows  us  to obtain a qualitative description of the local and global dynamics of
cosmological scenarios  independent of the initial conditions and the
specific evolution of the universe. Furthermore, one can find asymptotic
solutions and the corresponding  theoretical values  to compare with the observable ones.  Examples of such quantities are the DE
and total equation-of-state parameters, the deceleration parameter,
the density parameters for the different species, etc. These observables  allow to classify the cosmological solutions. In this regard, we can follow  th  reference \cite{Ellis},   the first book related to modern dynamical systems theory to both cosmological models and observations. 

In order to perform the stability analysis of a given cosmological scenario,
one first transforms it to its autonomous form $\label{eomscol}
\textbf{X}'=\textbf{f(X)}$
\citep{Ellis,Ferreira:1997au,Copeland:1997et,Perko,Coley:2003mj,Copeland:2006wr,Chen:2008ft,Cotsakis:2013zha,Giambo:2009byn},
where $\textbf{X}$ is a column vector containing some auxiliary variables and primes denote derivative
with respect to a time variable (conveniently chosen). Then, one extracts the  critical points
$\bf{X_c}$  by imposing the condition  $\bf{X}'=0$ and, in order to determine
their stability properties, one expands around them with $\textbf{U}$ the
column vector of the perturbations of the variables. Therefore,
for each critical point the perturbation equations are expanded to first
order as $\label{perturbation} \textbf{U}'={\bf{Q}}\cdot
\textbf{U}$, with the matrix ${\bf {Q}}$ containing the coefficients of the
perturbation equations. The eigenvalues of ${\bf {Q}}$ determine the type and
stability of the specific critical point.

\subsection{Stability analysis of Model I} \label{sec:SAMatter}

To start the dynamical analysis for the Barrow cosmology (\S \ref{sec:cosmo}), we use the dynamical variables $( \Omega_{m}, \Omega_{\Lambda})$ defined in \eqref{New_Vars_1}, say,  
\begin{equation}
    \Omega_{m}=  \frac{8 \pi  G \rho_{m}}{3 H^2}, \quad \Omega_{\Lambda}=\frac{\Lambda }{3 H^2}. 
\end{equation}
As we commented before, the normalized Friedmann equation
\eqref{normalized:Friedman} is then transformed to \begin{equation}
    \label{normalized:Friedman_2}
 \Omega_{\Lambda}+\Omega_m=\frac{\beta  (\Delta +2)
   H^{-\Delta }}{2- \Delta}.  
\end{equation} 
Notice that by substituting $\beta=1, \Delta=0$, in Eq.  \eqref{normalized:Friedman}, we obtain the usual relation in FRW cosmology.
For $\beta\neq 1, \Delta\neq 0$, by substituting 
Eqs. \eqref{ec:rhoDE} and
\eqref{ec:PDE} into Eq. \eqref{ec:raychaudhuri}, we obtain 
\begin{align}
    & 8 \pi  G \rho_{m}+\beta  (\Delta +2) H^{-\Delta } \dot{H}=0,\\
    &\implies 3 H^2 \Omega_m +\beta  (\Delta +2)  H^{-\Delta } \dot{H}=0, \\
   & \implies 3 H^2 \Omega_m + (2- \Delta)\left( \Omega_{\Lambda}+\Omega_m\right)\dot{H}=0.
\end{align}
Then, for $H \neq 0$, $\beta\neq 1$, and $\Delta\neq 0$, the deceleration parameter (Eq. \eqref{eq:qz}) results
\begin{equation}
\label{general_q}
q= -1 +\frac{3 \Omega_m}{(2-\Delta)(\Omega_\Lambda +\Omega_m)},  
\end{equation}
for  $\Lambda$CDM  ($\beta= 1, \Delta=0$) we obtain the usual relation $
q= -1 +\frac{3}{2} \Omega_{m}$.

In general, $H(t)$ satisfies the differential equation 
\begin{equation}
\dot {H}=\frac{\Lambda  H^{\Delta }}{\beta  (\Delta +2)}+\frac{3
   H^2}{\Delta -2}.
\end{equation}
Furthermore, we have
\begin{equation}
\rho _{m}= -\frac{3 \beta  (\Delta +2) H^{2-\Delta }}{8 \pi 
   (\Delta -2) G}-\frac{\Lambda }{8 \pi  G}. 
\end{equation}
  Finally, we obtain the dynamical system
 \begin{align}
     & \Omega_\Lambda'= 2 (q+1) \Omega_\Lambda,\quad 
      \Omega_m'= (2 q-1) \Omega_m,
 \end{align}
where the prime means derivative with respect $\tau= \ln(a)$, and $q$ is defined by \eqref{general_q}.
The main difference with the  $\Lambda$CDM model is that the term 
 $\frac{\beta  (\Delta +2)
   H^{-\Delta }}{2- \Delta}$  in Eq. \eqref{normalized:Friedman_2}  is  unbounded as $H \rightarrow 0$, resulting in unbounded $\Omega_\Lambda,
\Omega_m$. 
The equilibrium points in the finite part of the phase space are
\begin{enumerate}
    \item   the  line $A(\Omega_\Lambda):$ $\Omega_m=0$, $\Omega_\Lambda=\text{arbitrary}$, for  $\Delta=\text{arbitrary}$, with eigensystem $\left(
\begin{array}{cc}
 0 & -3 \\
 \{1,0\} & \left\{\frac{2}{\Delta -2},1\right\} \\
\end{array}
\right)$; and 
    \item \ the line $B(\Omega_m):$ $\Omega_m=\text{arbitrary}, \Omega_\Lambda=0$, for $\Delta=0$, with eigensystem
    $\left(
\begin{array}{cc}
 3 & 0 \\
 \{-1,1\} & \{0,1\} \\
\end{array}
\right)$.
\end{enumerate} 
The line of points  $B(\Omega_m)$ exists only for $\Delta=0$.
All these lines of equilibrium points are normally hyperbolic   because  the tangent vector at a given point of each line is parallel to the corresponding eigenvector associated to the zero eigenvalue. This implies that the stability conditions can be inferred from the eigenvalues with non-zero real parts \cite{aulbach1984continuous}.  Therefore, the line $A(\Omega_\Lambda)$ is the attractor of the system, representing de Sitter solutions. For $\Delta=0$, the line $B(\Omega_m)$ contains the past attractors, which represents  matter dominated solutions.    
   
\subsubsection{Global dynamical systems formulation}

In this section we define the compact variables (assuming $H \geq 0, H_0>0$) based on the approach by  \cite{Alho:2014fha}: 
\begin{equation}
T=  \frac{H_0}{H_0+H},  
\end{equation}
along with the angular variable
\begin{equation}
    \theta= \tan^{-1}\left(\sqrt{\frac{\Omega_m}{\Omega_\Lambda}}\right)= \tan ^{-1}\left(\sqrt{\frac{8 \pi G \rho _{m}}{\Lambda
   }}\right),
\end{equation}
with inverse
\begin{equation}
 H= \frac{H_0(1-T)}{T}, \quad \rho_{m}=\frac{\Lambda  \tan
   ^2(\theta )}{8 \pi  G}.
\end{equation}
We obtain the dynamical system  
\begin{align}
& \frac{d T}{d \bar{\tau}}=-\frac{3 (1-T)^2 T}{\Delta -2}   -\frac{\Lambda  T^3 {H_0}^{\Delta -2} \left(\frac{1}{T}-1\right)^{\Delta }}{\beta  (\Delta +2)}, \nonumber\\
& \frac{d \theta}{d \bar{\tau}}=-\frac{3}{4} (1-T) \sin (2 \theta ), \label{eq:40}
\end{align}
 \begin{table}[t!]
    \centering
    \begin{tabular}{|c|c|c|c|}
    \hline
    Label & Coordinates & Eigenvalues & Stability \\
    \hline
$dS_+$& $\left\{\theta = 2 n \pi\right\}$ & $\left\{-\frac{3}{2},0\right\}$ & sink \\
$dS_-$& $\left\{ \theta = (2n+1) \pi\right\}$ & $\left\{-\frac{3}{2},0\right\}$ & sink \\
$M_{-}^{(0)}$& $\left\{T=0,\theta= (4n-1)\frac{\pi}{2}\right\}$ & $\left\{\frac{3}{2-\Delta},\frac{3}{2}\right\}$ & source \\
$M_{+}^{(0)}$ & $\left\{T=0,\theta=(4n+1)\frac{\pi}{2}\right\}$ & $\left\{\frac{3}{2-\Delta},\frac{3}{2}\right\}$ & source\\
$M_{-}^{(1)}$&  $\left\{T= 1,\theta = (4n-1)\frac{\pi}{2}\right\}$ & $\left\{-\frac{3}{2-\Delta},\frac{3}{2}\right\}$ & saddle \\
$M_{+}^{(1)}$ & $\left\{T=1,\theta = (4n+1)\frac{\pi}{2}\right\}$ & $\left\{-\frac{3}{2-\Delta},\frac{3}{2}\right\}$ & saddle\\
\hline
\end{tabular}
    \caption{Equilibrium points/lines of  system  \eqref{eq:46}.}
    \label{tab:my_label2}
\end{table}
where for a function $f \in \{T, \theta\}$ we have introduced the new derivative 
 $$\frac{d f}{d \bar{\tau}}=\frac{1}{(H_0 + H)} \frac{d f}{dt},$$
 which allows for a global dynamical system analysis. 
 
 On the other hand, 
 the equation \eqref{normalized:Friedman_2} leads to
\begin{equation}
\frac{\Lambda  T^2 \sec ^2(\theta )}{3 {H_0}^2 (1-T)^2}+\frac{\beta  (\Delta +2) \left(H_0 \left(\frac{1}{T}-1\right)\right)^{-\Delta }}{\Delta -2}=0,    
\end{equation} 
that can be solved globally for $\beta$, and we end up with the system
\begin{align}
& \frac{d T}{d \bar{\tau}}=\frac{3 (1-T)^2 T \sin ^2(\theta )}{2-\Delta}, \nonumber \\
& \frac{d \theta}{d \bar{\tau}}= -\frac{3}{4} (1-T) \sin (2 \theta ),\label{eq:44}
\end{align} 
defined in the finite cylinder $\mathbf{S}$ with boundaries $T=0$ and $T=1$,
where the power-law dependence in the first equation \eqref{eq:40} is eliminated. Using the logarithmic variable $\tau= \ln (a)$, we obtain the complementary system 
 \begin{align}
& \frac{d T}{d{\tau}}=\frac{3 (1-T) T \sin ^2(\theta )}{2-\Delta}, \nonumber \\
& \frac{d \theta}{d {\tau}}=-\frac{3}{4} \sin (2 \theta ). \label{eq:46}
\end{align}
The deceleration parameter is now written as
\begin{equation}
    q=-1+\frac{3\Omega_{m}}{(2-\Delta ) \Omega_\Lambda 
   \left(\frac{\Omega_{m}}{\Omega_\Lambda }+1\right)}
  = -1+\frac{3 \sin ^2(\theta )}{2-\Delta }.
\end{equation}
From the first equation in \eqref{eq:46}, $T$ is a monotonically increasing function on $\mathbf{S}$. As a consequence, all orbits originate from the invariant subset $T=0$ (which contains the $\alpha$-limit), which is classically related to the initial singularity with $H \rightarrow \infty$, and ends on the invariant boundary subset $T=1$, which corresponds asymptotically to $H=0$. 

We have the relations for the fractional energy densities: 
\begin{align}
  & \Omega_{m}=\frac{\beta  (\Delta +2) H_0^{1-\Delta } \left(\frac{1}{T}-1\right)^{1-\Delta }}{2 (2-\Delta)}- \frac{\Lambda  T}{6 H_0 (1-T)},\\
   & \Omega_\Lambda =\frac{\Lambda  T}{6 H_0(1-T)}.
\end{align}

The system  \eqref{eq:46} admits the equilibrium points summarized in Table \ref{tab:my_label2}:
\begin{figure*}[t!]
    \centering
    \includegraphics[width=1.0\textwidth]{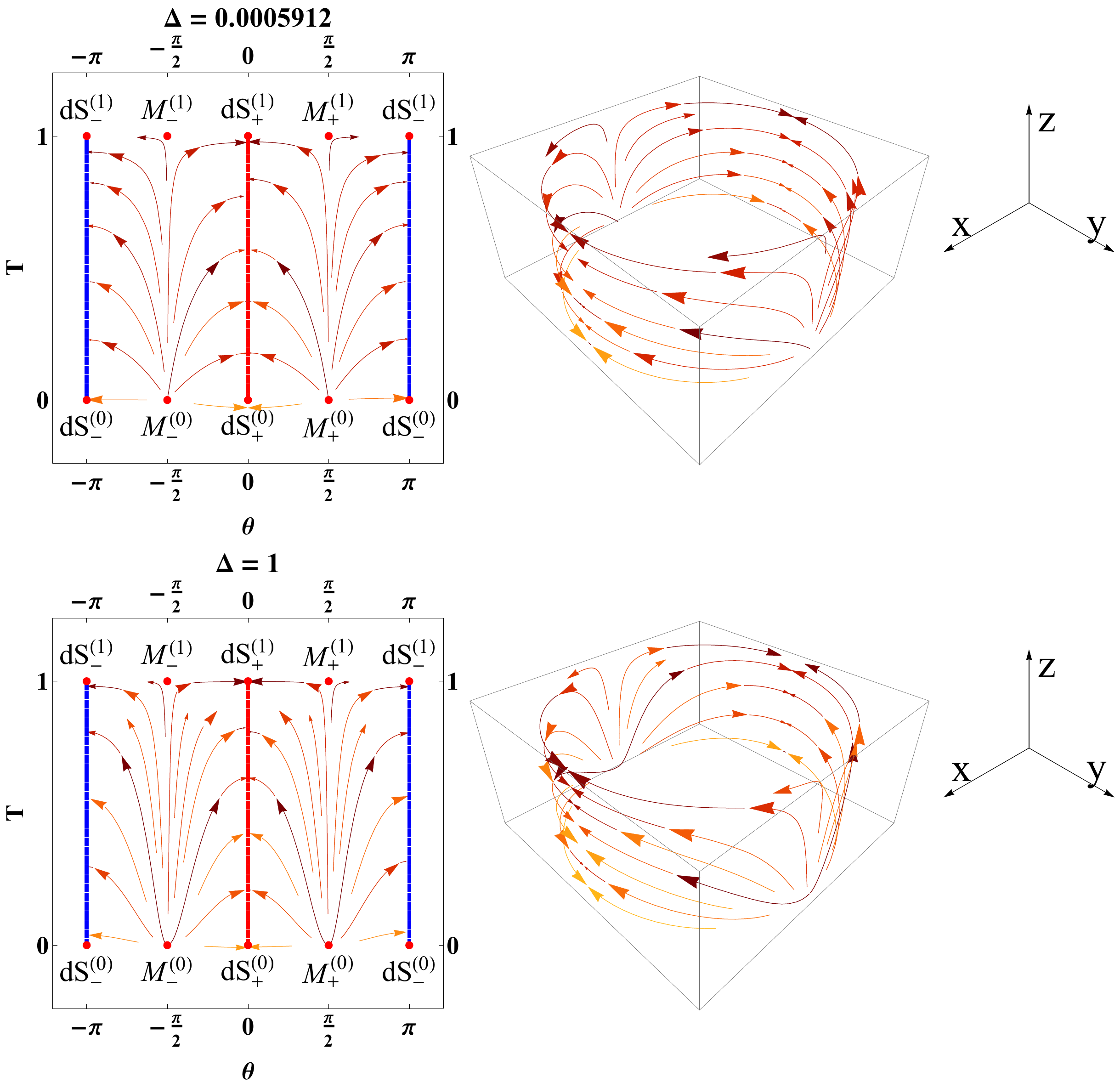}
    \caption{Unwrapped solution space (left panel) and projection over the cylinder $\mathbf{S}$ defined in Cartesian coordinates $(x,y,z)$  by \eqref{cylinder} (right panel) of the  solution space of system   \eqref{eq:44} for $\Delta=5.912\times 10^{-4}$, $\bar{\beta}=0.920$ (top panel) and $\Delta=1$, $\bar{\beta}=1$ (bottom panel).}
    \label{fig:my_label}
\end{figure*}
\begin{enumerate}
    \item $M_{\pm}^{(0)}: \theta = (4n \pm 1)\frac{\pi}{2}, T=0$, represented in Fig. \ref{fig:my_label}; 
      \item $M_{\pm}^{(1)}: \theta = (4n \pm 1)\frac{\pi}{2}, T=1$, represented in Fig. \ref{fig:my_label}; 
      \item $d S_+: \theta=2 n \pi, T=\text{arbitrary}$, with representatives $d S_+^{(0)}$ and  $d S_+^{(1)}$ respectively, denoted in Fig. \ref{fig:my_label} by a red line;
      \item $d S_-: \theta=(2 n +1) \pi, T=\text{arbitrary}$,  with representatives $d S_-^{(0)}$ and  $d S_-^{(1)}$ respectively, denoted in Fig. \ref{fig:my_label} by a blue line;
\end{enumerate}
where $n$ is an integer. There are two equivalent (due to the discrete symmetry)
hyperbolic fixed points $M_{\pm}$ for which $q =-1+\frac{3 }{2-\Delta }$, i.e. they are associated with dust fluid for $\Delta=0$, and two equivalent fixed points $dS_{\pm}$ for which $q = -1$, which therefore correspond to a de Sitter state.

For a representation of the flow of \eqref{eq:46}, we integrate it in the variables $T, \theta$ and project in a compact set
using the ``cylinder-adapted'' coordinates 
\begin{equation}
\label{cylinder}
 \mathbf{S}: \begin{cases}
x = \cos \theta,\\
y = \sin \theta,\\
z = T, 
\end{cases}
\end{equation}
with $0\leq T\leq 1, \theta_1 \in [-\pi, \pi]$,
with inverse 
\begin{equation}
\begin{cases}
\theta = \tan^{-1} \left(\frac{y}{x}\right),\\
T =z. 
\end{cases}
\end{equation}
Also, we  present the whole evolution in the space $(\Omega_m,  T)$ through the transform 
$(T, \theta)\mapsto (\Omega_m, T)$ where 
\begin{align}
 \Omega_m & = \left[\left(\frac{2+\Delta}{2-\Delta}\right)\frac{1}{\bar{\beta}}-\Omega_{m0}\right] \frac{T^2}{(1-T)^2 } \tan^2(\theta).
\end{align}
The value $T=1/2$ corresponds to current time. In these variables we have the system 
\begin{align}
  & \frac{d\Omega_m}{d\tau}=  \frac{3 \Omega _m \left(\bar{\beta } \left((\Delta -2) T^2 \Omega _{m0}-\Delta  (1-T)^2
   \Omega _m\right)+(\Delta +2) T^2\right)}{(\Delta -2) \bar{\beta } \left((1-T)^2 \Omega _m-T^2 \Omega
   _{m0}\right)-(\Delta +2) T^2}, \nonumber\\
  & \frac{d T}{d\tau}= \frac{3 (T-1)^3 T \bar{\beta } \Omega _m}{(\Delta -2) \bar{\beta
   } \left((1-T)^2 \Omega _m-T^2 \Omega _{m0}\right)-(\Delta +2) T^2} \label{syst4.23}.
\end{align}
In Figure \ref{fig:my_label} are represented the streamlines of the flow of \eqref{eq:44} onto the unwrapped solution space (left panel) and its projection over the cylinder $\mathbf{S}$ (right panel)  for two values of $\Delta$ one  obtained from the joint constraint $\Delta=5.912\times 10^{-4}$, and  the other one is the extreme value $\Delta=1$. The phase space is qualitatively the same as for system  \eqref{eq:46}. Firstly, observe that systems   \eqref{eq:44} and \eqref{eq:46} are independent of $\bar{\beta}$. Therefore, the parameter $\bar{\beta}$ is  dynamically irrelevant.  The results summarized in points 1-4 before are confirmed in  Figure \ref{fig:my_label}. That is, the matter dominated solutions  $M_{\pm}^{(0)}: \theta = (4n \pm 1)\frac{\pi}{2}, T=0 \;  (H\rightarrow \infty)$ are past attractors. The matter dominated solutions  $M_{\pm}^{(1)}: \theta = (4n \pm 1)\frac{\pi}{2},  T=1 \;(H\rightarrow 0)$ are saddle and the attractor is the line of equilibrium points connecting  $d S_-^{(0)}$ and  $d S_-^{(1)}$ which are de Sitter solutions. 

\begin{figure*}[t]
    \centering
    \subfigure[]{\includegraphics[scale = 0.45]{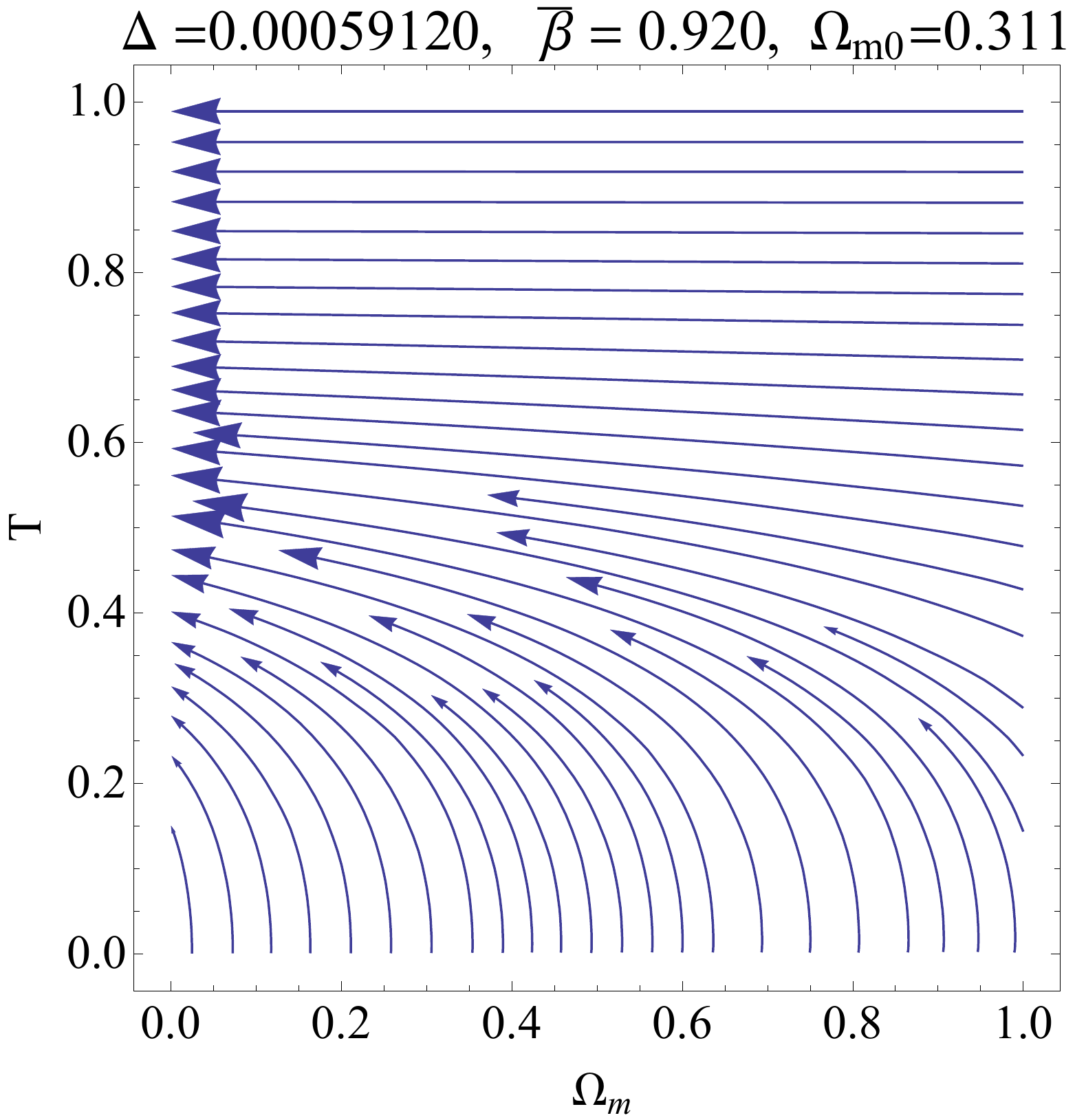}} \hspace{1cm}
     \subfigure[]{\includegraphics[scale = 0.45]{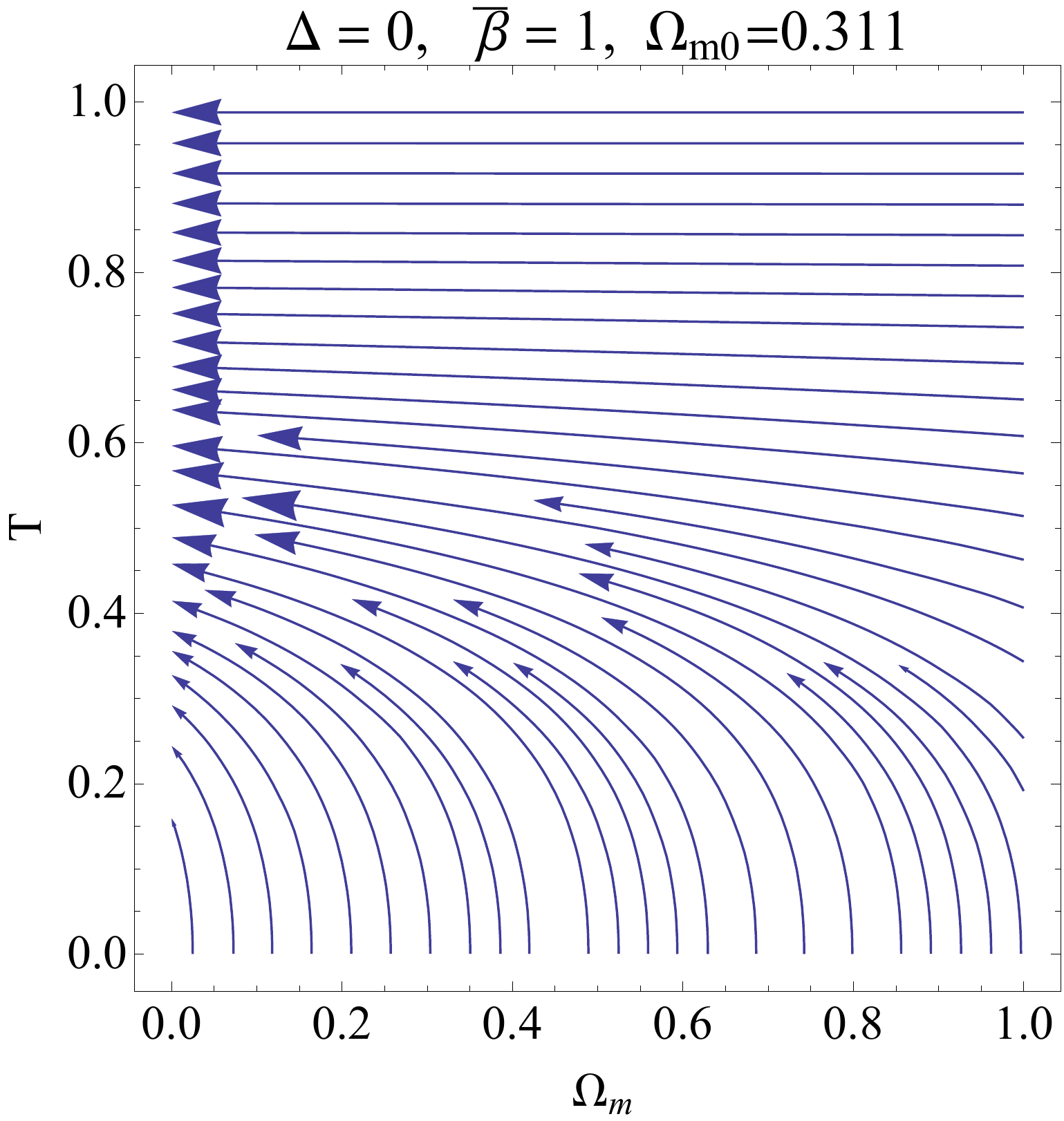}}
    \caption{\label{Fig4} Streamlines of the flow of \eqref{syst4.23} for (a) $\Delta=5.912\times 10^{-4}, \bar{\beta}= 0.920, \Omega_{m0}=0.311$ obtained from the joint constraints and (b) $\Delta=0, \beta= 1, \Omega_{m0}=0.311$ obtained from $\Lambda$CDM model. }
\end{figure*}

In Figure \ref{Fig4} streamlines of the flow of \eqref{syst4.23} are presented. We select the parameters (a) $\Delta=5.912\times 10^{-4}, \bar{\beta}= 0.920, \Omega_{m0}=0.311$ obtained from the joint constraints and (b) $\Delta=0, \beta= 1, \Omega_{m0}=0.311$ for $\Lambda$CDM model. These plots confirms that the late time attractor is the line of de Sitter points 
$d S_+$ and $d S_-$.

Complementary, we define an unbounded variable  $\tilde{T}$ and keep $\theta$: 
\begin{equation}
    \tilde{T}=\frac{H_0}{H}, \quad \theta=  \tan ^{-1}\left(\sqrt{\frac{8 \pi G \rho _{m}}{\Lambda
   }}\right),
\end{equation}
and  using the logarithmic variable $\tau= \ln (a)=-\ln(1+z)$, we obtain the complementary system 
 \begin{align}
 \label{eq:51}
& \frac{d  \tilde{T}}{d{\tau}}=\frac{3  \tilde{T} \sin ^2(\theta )}{2-\Delta}, \nonumber \\
& \frac{d \theta}{d {\tau}}=-\frac{3}{4} \sin (2 \theta ).
\end{align}
Setting $a=1$ for the current universe, and considering the initial conditions: 
\begin{equation}
    \theta(0)= \tan ^{-1}\left(\sqrt{\frac{\Omega_{m0}}{\Omega_{\Lambda 0}}}\right), \quad \tilde{T}(0)=1,
\end{equation}
the system \eqref{eq:51} is integrated to obtain 
\begin{align}
    & \theta (\tau )= \tan ^{-1}\left(e^{-3 \tau /2} \sqrt{\frac{\Omega_{m0}}{\Omega_{\Lambda 0}}}\right),\\
   & \tilde{T}(\tau )=  e^{-\frac{3 \tau }{\Delta -2}} \left(\frac{\Omega_{\Lambda 0}+\Omega_{m0}}{\Omega_{\Lambda 0}}\right)^{\frac{1}{2-\Delta }} \left(e^{3 \tau }+\frac{\Omega_{m0}}{\Omega_{\Lambda 0}}\right)^{\frac{1}{\Delta -2}},
\end{align}
Finally, we get the exact evolution of $H$ and $\rho_m$: 
\begin{align}
 & H= H_0
 e^{-\frac{3 \tau }{2-\Delta}} \left(\frac{\Omega_{\Lambda 0}}{\Omega_{\Lambda 0}+\Omega_{m0}}\right)^{\frac{1}{2-\Delta }} \left(e^{3 \tau }+\frac{\Omega_{m0}}{\Omega_{\Lambda 0}}\right)^{\frac{1}{2-\Delta}} \nonumber
 \\
 & = H_0  \left(\frac{ \Omega_{m0}(z+1)^3+\Omega_{\Lambda 0}}{\Omega_{\Lambda 0}+\Omega_{m0}}\right)^{\frac{1}{2-\Delta }},\label{eqH4.27}
 \\
& \rho_{m}=\frac{\Lambda  \tan
   ^2(\theta )}{8 \pi  G}= \frac{\Lambda  e^{-3 \tau } \Omega_{m0}}{8 \pi  G \Omega_{\Lambda 0}}=\frac{3 H_0^2\Omega_{m0}}{8 \pi G} (z+1)^3.
\end{align}
Eq. \eqref{eqH4.27} can be deduced from Eq. \eqref{eq:Ez} after the substitution of $\left(\frac{2+\Delta}{2-\Delta}\right)\frac{1}{\bar{\beta}}$ from Eq. \eqref{eq2.15}. 

These expressions are used to obtain the fractional energy densities corresponding to matter, to an effective  a $\Lambda$-like source, and to the effective dark energy as follows: 
\begin{equation}
    \Omega_m(z)= \frac{8 \pi G \rho_m}{3 H^2}=   \Omega_{m0}   (z+1)^3 \left(\frac{ \Omega_{m0}(z+1)^3+\Omega_{\Lambda 0}}{\Omega_{\Lambda 0}+\Omega_{m0}}\right)^{-\frac{2}{2-\Delta }},
    \label{eq:om_mat}
\end{equation}
and 
\begin{equation}
    \Omega_\Lambda(z)= \frac{\Lambda}{3 H^2}= \Omega_{\Lambda 0}\left(\frac{ \Omega_{m0}(z+1)^3+\Omega_{\Lambda 0}}{\Omega_{\Lambda 0}+\Omega_{m0}}\right)^{-\frac{2}{2-\Delta }}, \quad  \Omega_{\Lambda 0}= \frac{\Lambda}{3 H_0^2}.
    \label{eq:ol_mat}
\end{equation}

From  Eq. \eqref{ec:rhoDE} we   infer 
 \begin{align} 
\label{ec:omegaDE}
&\Omega_{DE}(z)=\frac{8\pi G \rho_{DE}}{3 H^2}= \frac{\Lambda}{3H^2}+ \left[1-\frac{ \beta (\Delta+2)}{2-\Delta} 
H^{-\Delta}
\right]\nonumber \\
& = 1+ \Omega_\Lambda - \frac{ \beta (\Delta+2)}{2-\Delta} 
H_0^{-\Delta} \left(\frac{ \Omega_{m0}(z+1)^3+\Omega_{\Lambda 0}}{\Omega_{\Lambda 0}+\Omega_{m0}}\right)^{-\frac{\Delta}{2-\Delta }} \nonumber\\
 & = 1+ \Omega_{\Lambda 0}\left(\frac{ \Omega_{m0}(z+1)^3+\Omega_{\Lambda 0}}{\Omega_{\Lambda 0}+\Omega_{m0}}\right)^{-\frac{2}{2-\Delta }} - \frac{ (\Delta+2)}{(2-\Delta)} \frac{1}{\bar{\beta}}
\left(\frac{ \Omega_{m0}(z+1)^3+\Omega_{\Lambda 0}}{\Omega_{\Lambda 0}+\Omega_{m0}}\right)^{-\frac{\Delta}{2-\Delta }} \nonumber \\
&=  1+ \Omega_{\Lambda 0}\left(\frac{ \Omega_{m0}(z+1)^3+\Omega_{\Lambda 0}}{\Omega_{\Lambda 0}+\Omega_{m0}}\right)^{-\frac{2}{2-\Delta }} -\left(\Omega_{\Lambda0}+\Omega_{m0}\right) \left(\frac{ \Omega_{m0}(z+1)^3+\Omega_{\Lambda 0}}{\Omega_{\Lambda 0}+\Omega_{m0}}\right)^{-\frac{\Delta}{2-\Delta }}
 \nonumber\\
&=  1   -    \Omega_{m0} (1+z)^3 \left(\frac{\Omega_{m0}(1+z)^3 +  \Omega_{\Lambda 0}}{\Omega_{\Lambda 0}+\Omega_{m0}} \right)^{-\frac{2}{2-\Delta}}
 \end{align}
where we   use Eq. \eqref{eq2.15} to eliminate the term $\left(\frac{2+\Delta}{2-\Delta}\right)\frac{1}{\bar{\beta}}$,   apply  $-\frac{\Delta}{2-\Delta}= 1 -\frac{2}{2-\Delta}$,   and  algebraically  manipulate the equations to obtain $\Omega_{DE}=1- \Omega_m $. 
It is straightforward to    infer  Eqs. \eqref{eq:om_mat}, \eqref{eq:ol_mat} and \eqref{ec:omegaDE} in terms of the scale factor by replacing $z=(1/a) -1$.

\begin{figure}
\centering
\includegraphics[width=0.65\textwidth]{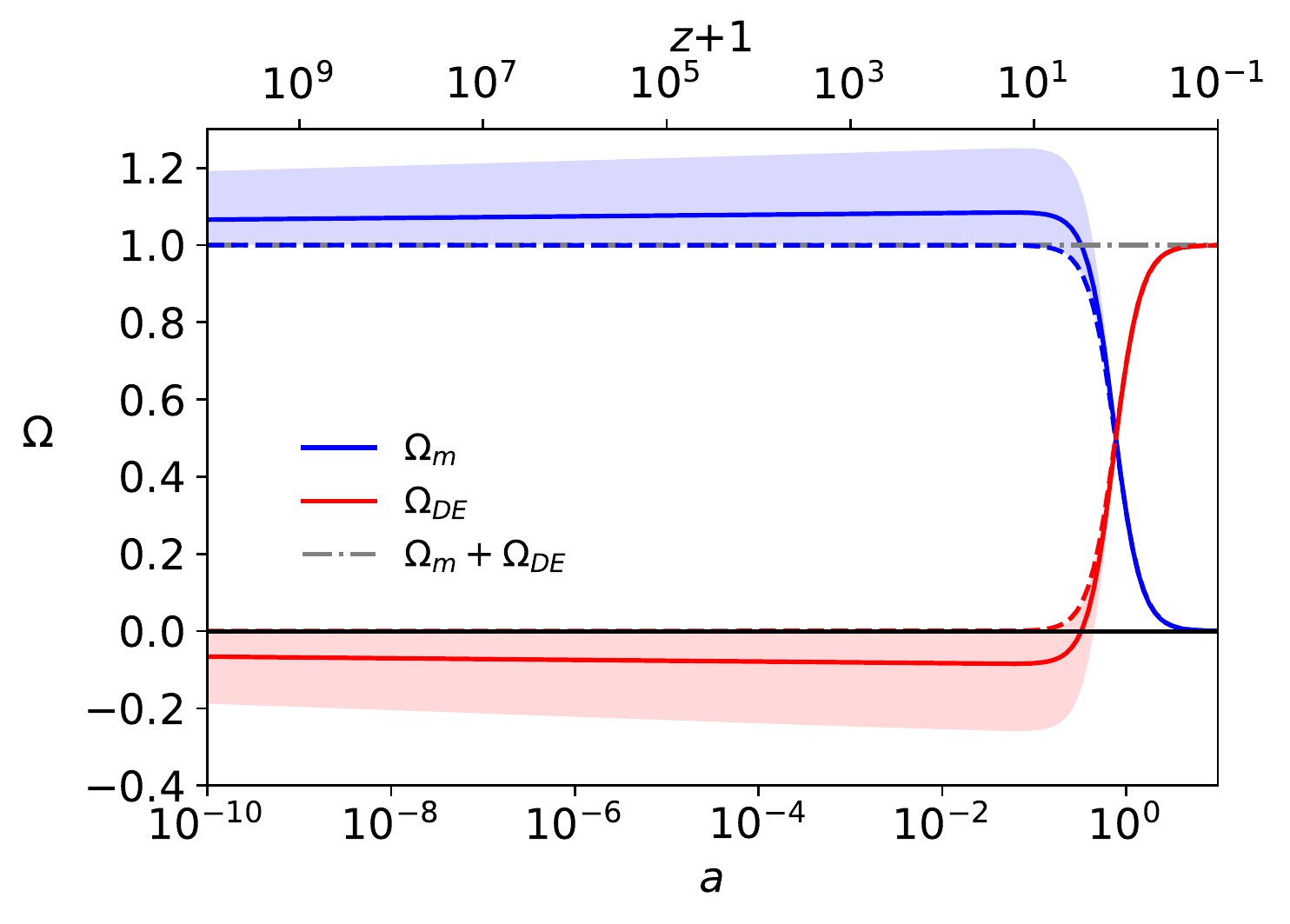}
\caption{Evolution of the density parameters for matter and effective dark energy in Barrow Cosmology using the joint constraints (solid lines) and the  standard model ($\Delta=0$, and $\bar \beta=1$, dashed lines). The shadow regions represent the $3\sigma$ confidence levels.}
\label{fig:omegas_mat}
\end{figure}

Figure \ref{fig:omegas_mat} shows the evolution of the density parameters $\Omega_m$ and $\Omega_{DE}$ as a function of the scale factor $a$ for two cases: the joint constraints for $h$, $\Omega_{m0}$, $\Delta$ and $\bar{\beta}$   presented in Table \ref{tab:bestfits} (solid lines) and the values for the standard model $\Delta=0$, $\bar\beta=1$ (dashed lines). The shadowed regions represent the $3\sigma$ confidence levels. It is worthy to note that there are values of the energy densities which satisfy $\Omega_m>1$ and $\Omega_{DE}<0$, however, these values are close to $\Lambda$CDM lines within the $3\sigma$ error propagation. 

\subsection{Stability analysis of Model II} \label{sec:SB}
We start our stability study of the dynamical variables \eqref{New_Vars_1} and \eqref{varOmega_r}, say, \begin{equation}
    \Omega_{m}=  \frac{8 \pi  G \rho_{m}}{3 H^2}, \quad \Omega_{r}=  \frac{8 \pi  G \rho_{r}}{3 H^2}, \quad \Omega_{\Lambda}=\frac{\Lambda }{3 H^2}, 
\end{equation} related by \eqref{normalized:Friedman2_2}. 
The evolution equations are now given by 
\begin{align}
& \Omega_\Lambda' = 2 (q+1) \Omega_\Lambda, \nonumber  \\
& \Omega_m'= (2 q-1) \Omega_{m}, \nonumber \\
& \Omega_r'= 2 (q-1) \Omega_{r},
\end{align}
where the prime means derivative with respect to $\tau= \ln(a)$, and $q$ is defined by \eqref{general_q-rad} 
\begin{equation}
\label{general_q-rad}
q=-1+\frac{3 \Omega_m+4 \Omega_r}{(2-\Delta) (\Omega_\Lambda +\Omega_m+\Omega_r)}. 
\end{equation}
The equilibrium points in the finite part of the phase space are the lines of equilibrium points: 
\begin{enumerate}
    \item the line $A(\Omega_\Lambda):$ $\Omega_m=0, \Omega_r$, $\Omega_\Lambda=\text{arbitrary}$, for  $\Delta=\text{arbitrary}$, with eigensystem $\left(
\begin{array}{ccc}
 0 & -4 & -3 \\
 \{1,0,0\} & \left\{\frac{2}{\Delta -2},0,1\right\} & \left\{\frac{2}{\Delta -2},1,0\right\} \\
\end{array}
\right)$;    
    \item the line $B(\Omega_m):$ $\Omega_m=\text{arbitrary}, \Omega_{r}=0, \Omega_\Lambda=0$, for $\Delta=0$, with eigensystem
    $\left(
\begin{array}{ccc}
 3 & -1 & 0 \\
 \{-1,1,0\} & \{0,-1,1\} & \{0,1,0\} \\
\end{array}
\right)$; and 
    \item  the line $C(\Omega_r):$ $\Omega_{m}=0, \Omega_r=\text{arbitrary}, \Omega_\Lambda=0$, for $\Delta=0$, with eigensystem
    $\left(
\begin{array}{ccc}
 4 & 1 & 0 \\
 \{-1,0,1\} & \{0,-1,1\} & \{0,0,1\} \\
\end{array}
\right)$.
\end{enumerate} 
The last   two  lines of equilibrium points, $B(\Omega_m)$ and $C(\Omega_r)$, exist only for $\Delta=0$.
All   three lines are normally hyperbolic   because the tangent vector at a given point of each line is parallel to the corresponding eigenvector associated to the zero eigenvalue. This implies that the stability conditions can be inferred from the eigenvalues with non-zero real parts \cite{aulbach1984continuous}.  Therefore, the line $A(\Omega_\Lambda)$ is the attractor of the system, representing de Sitter solutions. For $\Delta=0$, the line $B(\Omega_m)$ is a saddle, representing matter dominated solutions, and  the line $C(\Omega_r)$ contains the past attractors, which represents radiation dominated solutions.

\subsubsection{Global dynamical systems formulation}

In this section we define the compact   variable  (assuming $H \geq 0, {H_0}>0$) based on the approach by  \cite{Alho:2014fha}: 
\begin{equation}
T=  \frac{{H_0}}{{H_0}+H},  
\end{equation} 
along with the angular   ones 
\begin{equation}
    \theta_1=  \tan ^{-1}\left(\sqrt{\frac{8 \pi G \rho _{m}}{\Lambda
   }}\right),  \quad   \theta_2=  \tan ^{-1}\left( \sqrt{\frac{8 \pi G \rho _{r}}{\Lambda
   }}\right),
\end{equation}
with inverse
\begin{equation}
 H= \frac{H_0(1-T)}{T}, \quad \rho_{m}=\frac{\Lambda  \tan
   ^2(\theta_1)}{8 \pi  G}, \quad  \rho_{r}=\frac{\Lambda  \tan
   ^2(\theta_2)}{8 \pi  G}.
\end{equation}
Furthermore, we have
\begin{small}
\begin{equation}
    \beta = \frac{(2-\Delta) \Lambda  T^2 \left(\tan ^2(\theta _{1})+\tan ^2( \theta_{2})+1\right) \left({H_0} \left(\frac{1}{T}-1\right)\right)^{\Delta }}{3 (\Delta +2){H_0}^2 (1-T)^2},
\end{equation}
\end{small}
and
\begin{equation}
  q=-1+  \frac{3 \tan ^2(\theta_1)+4 \tan ^2(\theta_2)}{(2-\Delta ) \left(\tan ^2(\theta_1)+\tan ^2(\theta_2)+1\right)}
\end{equation}

We obtain the dynamical system  
\begin{align}
& \frac{d T}{d \bar{\tau}}=\frac{(1-T)^2 T \left(3 \tan ^2(\theta_1)+4 \tan ^2(\theta_2)\right)}{(2-\Delta) \left(\tan ^2(\theta_1)+\tan ^2(\theta_2)+1\right)}, \nonumber\\
& \frac{d \theta_1}{d \bar{\tau}}=-\frac{3}{4} (1-T) \sin (2 \theta_1 ), \nonumber\\
& \frac{d \theta_2}{d \bar{\tau}}=- (1-T) \sin (2 \theta_2), \label{eq:82}
\end{align}
where for any function $f \in \{T, \theta_1, \theta_2\}$ we have introduced the new derivative 
 $$\frac{d f}{d \bar{\tau}}=\frac{1}{({H_0} + H)} \frac{d f}{dt},$$
 which allows for a global dynamical system analysis. 
 
Using the logarithmic variable $\tau= \ln (a)$, we  get the complementary system 
\begin{align}
& \frac{d T}{d {\tau}}=\frac{(1-T) T \left(3 \tan ^2(\theta_1)+4 \tan ^2(\theta_2)\right)}{(2-\Delta) \left(\tan ^2(\theta_1)+\tan ^2(\theta_2)+1\right)}, \nonumber\\
& \frac{d \theta_1}{d {\tau}}=-\frac{3}{4} \sin (2 \theta_1 ), \nonumber\\
& \frac{d \theta_2}{d {\tau}}=- \sin (2 \theta_2). \label{eq:83}
\end{align}

There exist three  classes of equilibrium points/lines:  $dS$,  $R$, and $M$. The deceleration parameter $q$ evaluated at the lines $dS$ is 
$q=-1$,    thus,  it denotes the de Sitter solutions. Evaluating $q$ at the points $R$ and $M$ we have 
$q=-1+  \frac{4}{(2-\Delta )}$ or $q =-1+\frac{3 }{2-\Delta }$, respectively; i.e.   for $\Delta=0$  they are associated   with radiation-dominated and with dust fluid  solutions, respectively.

\begin{table*}[t!]
    \centering
    \begin{tabular}{|c|c|c|c|}
    \hline
    Label & Coordinates & Eigenvalues & Stability \\
    \hline
$dS_{++}$ & $\{\theta_1= 2 n \pi ,\theta_2= 2 m \pi \}$ & $\left\{-\frac{3}{2},-2,0\right\}$  & stable \\ \hline
$dS_{+-}$ & $\{\theta_1= 2 n \pi ,\theta_2= (2m+1)\pi \}$& $\left\{-\frac{3}{2},-2,0\right\}$ & stable \\ \hline
$dS_{-+}$ & $\{\theta_1= (2n+1)\pi ,\theta_2= 2 m \pi \}$& $\left\{-\frac{3}{2},-2,0\right\}$ & stable \\ \hline
$dS_{--}$ & $\{\theta_1= (2n+1)\pi ,\theta_2= (2m+1)\pi \}$& $\left\{-\frac{3}{2},-2,0\right\}$ & stable \\ \hline
$R_{+\pm}^{(0)}$& $\left\{T= 0,\theta_1= 2 n \pi ,\theta_2= \frac{1}{2} (4 m \pm 1) \pi \right\}$  &  $\left\{-\frac{3}{2},2,\frac{4}{2-\Delta}\right\}$  & saddle \\ \hline
$R_{-\pm}^{(0)}$ & $\left\{T= 0,\theta_1= (2n+1)\pi ,\theta_2= \frac{1}{2} (4 m\pm 1) \pi \right\}$ & $\left\{-\frac{3}{2},2,\frac{4}{2-\Delta}\right\}$  & saddle \\ \hline
$R_{+\pm}^{(1)}$ & $\left\{T= 1,\theta_1= 2 n \pi ,\theta_2= \frac{1}{2} (4 m\pm 1) \pi \right\}$ & $\left\{-\frac{3}{2},2,-\frac{4}{2-\Delta}\right\}$   & saddle \\ \hline
$R_{-\pm}^{(1)}$ & $\left\{T= 1,\theta_1= (2n+1)\pi ,\theta_2= \frac{1}{2} (4 m \pm 1) \pi \right\}$ & $\left\{-\frac{3}{2},2,-\frac{4}{2-\Delta}\right\}$  & saddle \\ \hline
$M_{\pm +}^{(0)}$ & $\left\{T= 0,\theta_1= \frac{1}{2} (4 n\pm 1) \pi ,\theta_2= 2 m \pi \right\}$ & $\left\{\frac{3}{2},-2,\frac{4}{2-\Delta}\right\}$  & saddle \\ \hline
$M_{\pm -}^{(0)}$ & $\left\{T= 0,\theta_1= \frac{1}{2} (4 n\pm 1) \pi ,\theta_2= (2m+1)\pi \right\}$ & $\left\{\frac{3}{2},-2,\frac{4}{2-\Delta}\right\}$  & saddle \\ \hline
$M_{\pm +}^{(1)}$ & $\left\{T= 1,\theta_1= \frac{1}{2} (4 n \pm 1) \pi ,\theta_2= 2 m \pi \right\}$ & $\left\{\frac{3}{2},-2,-\frac{4}{2-\Delta}\right\}$   & saddle \\ \hline
$M_{\pm -}^{(1)}$ & $\left\{T= 1,\theta_1= \frac{1}{2} (4 n\pm 1) \pi ,\theta_2= (2m+1)\pi \right\}$ & $\left\{\frac{3}{2},-2,-\frac{4}{2-\Delta}\right\}$   & saddle \\ \hline
\end{tabular}
    \caption{Equilibrium points/lines of   system  \eqref{eq:83}.}
    \label{tab:my_label2B}
\end{table*}

The equilibrium points/lines of  system  \eqref{eq:83} are summarized in table \ref{tab:my_label2B}. The label $dS_{++}$ means that $\theta_1$ and $\theta_2$ are both even multiples of $\pi$; $dS_{+-}$  means that $\theta_1$ is an even multiple of $\pi$ and $\theta_2$ is an odd multiple of $\pi$, and so on. The left sign in kernel $R$ is $+$ if $\theta_1$ is an even multiple of $\pi$, and   $-$ if  it is odd  multiple of $\pi$. The right sign in kernel $R$ is $+$ if $\theta_2$ is co-terminal of $\frac{\pi}{2}$, and    $-$ if  it is co-terminal of  $-\frac{\pi}{2}$. For kernel $M$,  the left sign is $+$ if $\theta_1$ is co-terminal of $\frac{\pi}{2}$, and  $-$ if it is co-terminal of $-\frac{\pi}{2}$, whereas the right sign in kernel $M$ is $+$ if $\theta_2$ is an even multiple of $\pi$, and   $-$ if   it is  odd  multiple of $\pi$. For $M$, $R$- points, the upper indexes are ${(0)}$ or ${(1)}$ depending on whether $T=0$ or $T=1$.

As summarized in table \ref{tab:my_label2B}, we  found  three  classes of equilibrium points/lines:  
 \begin{enumerate}
     \item  the family $dS$, which comprises the lines of equilibrium points $dS_{++}$, $dS_{+-}$, $dS_{-+}$ and $dS_{--}$.  They    represent the de Sitter solutions  and are  stable. 
     \item  The family $R$, which  encompass the equilibrium points  $R_{+\pm}^{(0)}$, $R_{-\pm}^{(0)}$, $R_{+\pm}^{(1)}$ and $R_{-\pm}^{(1)}$. For $\Delta=0$, they  are associated to radiation-dominated solutions and are saddles. 
     \item  The family $M$, which    contains  the equilibrium points $M_{\pm +}^{(0)}$, $M_{\pm -}^{(0)}$, $M_{\pm +}^{(1)}$ and $M_{\pm -}^{(1)}$.  For $\Delta=0$, they  are associated with dust fluid solutions  and are saddles. 
 \end{enumerate}

Using $E:=\frac{H}{H_0}=\frac{1-T}{T}$ we obtain 
\begin{equation}
    E'=-E \left[\frac{3 \tan ^2(\theta_1)+4 \tan ^2(\theta_2)}{(2-\Delta ) \left(\tan ^2(\theta_1)+\sec ^2(\theta_2)\right)}\right].
\end{equation}
This implies that $E$ is a monotonic decreasing function. According to the monotonicity principle, the late time attractors satisfy $T=1$, whereas the early time attractors satisfy $T=0$. 
For a representation of the flow of \eqref{eq:83}, we integrate   in the variables $T, \theta_1, \theta_2$ and project in a compact set using the ``torus-adapted'' coordinates 
\begin{equation}
\label{torus}
 \mathbf{T}^1: \begin{cases}
x = \cos \theta_1 (2 + T \cos \theta_2),\\
y = \sin \theta_1  (2 + T \cos \theta_2),\\
z = T \sin \theta_2, 
\end{cases}
\end{equation}
with $0\leq T\leq 1, \theta_1, \theta_2 \in [-\pi, \pi]$,
with inverse 
\begin{equation}
\begin{cases}
\theta_1 = \tan^{-1} \left(\frac{y}{x}\right),\\
\theta_2 = \tan^{-1} \left(\frac{z}{\sqrt{x^2+y^2}-2}\right),\\
T =\sqrt{\left(\sqrt{x^2+y^2}-2\right)^2+z^2}. 
\end{cases}
\end{equation}

\begin{figure*}[t!]
    \centering
    \includegraphics[width=1.0\textwidth]{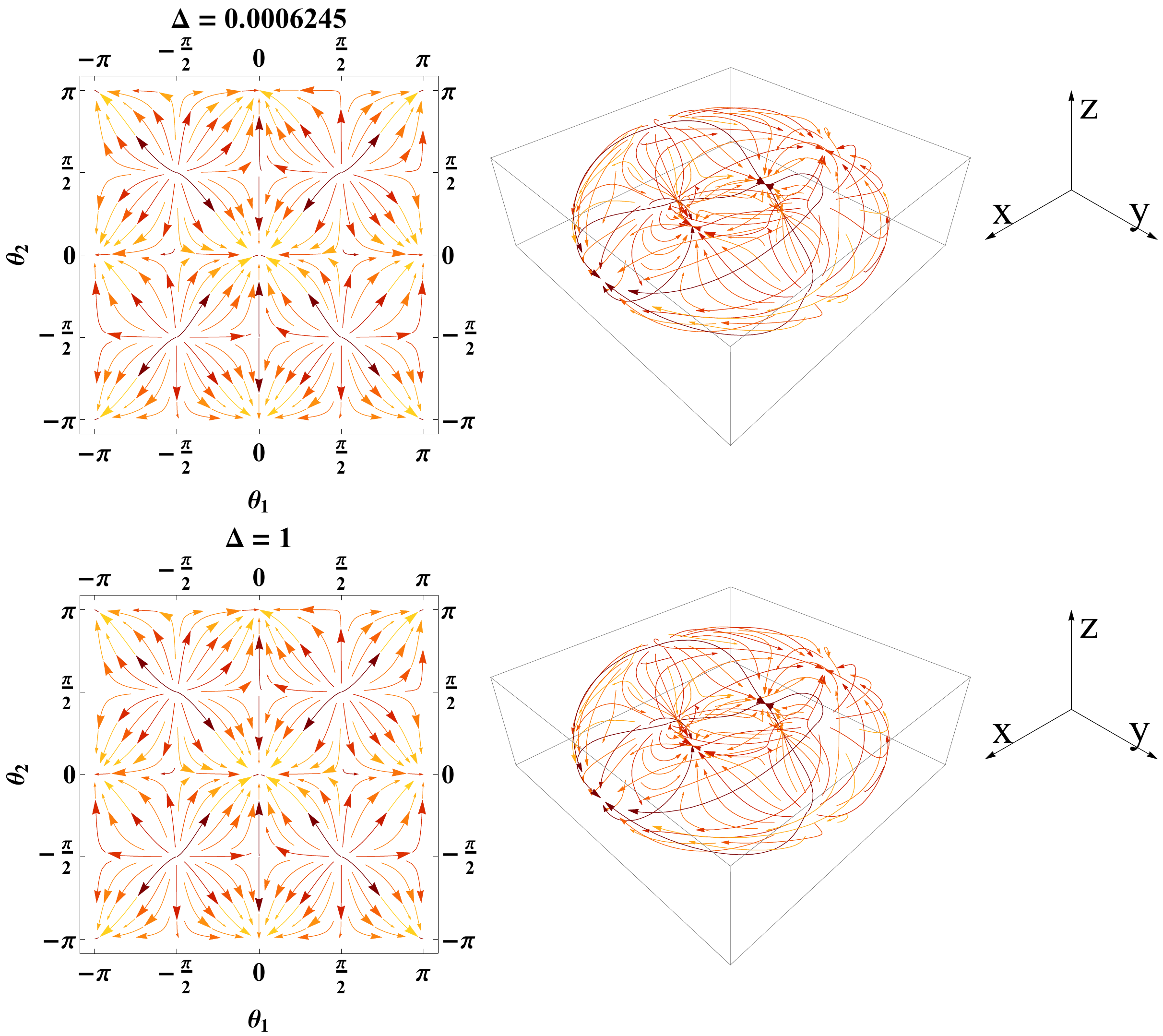}
    \caption{Unwrapped solution space (left panel) projected both   onto the plane $\theta_1, \theta_2$ and  the torus $\mathbf{T}^1$ given in Cartesian coordinates by \eqref{torus} (right panel) of the  solution space of system   \eqref{eq:83} (in the invariant set $T=1$) for  the cases  $\Delta=6.245\times 10^{-4}$   (top panels), and $\Delta=1$ (bottom panels). }
    \label{fig:my_label-3}
\end{figure*}
 In addition, we  present the whole evolution in the space $(\Omega_m, \Omega_r, T)$ through the transform 
$(T, \theta_1, \theta_2)\mapsto (\Omega_m,\Omega_r, T)$ where 
\begin{align}
\left( \Omega_m, \Omega_r\right) & = \left[\left(\frac{2+\Delta}{2-\Delta}\right)\frac{1}{\bar{\beta}}-\Omega_{m0}-\Omega_{r0}  \right] \frac{T^2}{(1-T)^2 } \left(\tan^2(\theta_1), \tan^2(\theta_2)\right).
\end{align}
The case $T=1/2$ corresponds to current time.  Using these variables we have the system 
\begin{small}
\begin{align}
&  \frac{d\Omega_m}{d\tau}=  -\Omega _m \left(\frac{2 (1-T)^2 \bar{\beta } \left(3 \Omega _m+4 \Omega _r\right)}{(\Delta -2) \bar{\beta } \left((1-T)^2 \Omega _m-T \left(T \left(\Omega _{m0}-\Omega _r+\Omega _{r0}\right)+2 \Omega
   _r\right)+\Omega _r\right)-(\Delta +2) T^2}+3\right),\nonumber 
  \\
&  \frac{d\Omega_r}{d\tau}=   -2 \Omega _r \left(\frac{(1-T)^2 \bar{\beta } \left(3 \Omega _m+4 \Omega _r\right)}{(\Delta -2) \bar{\beta } \left((1-T)^2 \Omega _m-T \left(T \left(\Omega
   _{m0}-\Omega _r+\Omega _{r0}\right)+2 \Omega _r\right)+\Omega _r\right)-(\Delta +2) T^2}+2\right), \nonumber \\
&  \frac{dT}{d\tau}=  -\frac{(1-T)^3 T \bar{\beta } \left(3 \Omega _m+4 \Omega _r\right)}{(\Delta -2) \bar{\beta } \left((1-T)^2 \Omega _m-T \left(T \left(\Omega _{m0}-\Omega _r+\Omega _{r0}\right)+2 \Omega _r\right)+\Omega
   _r\right)-(\Delta +2) T^2}. \label{syst4.49}
\end{align}
\end{small}

\begin{figure*}[t]
    \centering
    \subfigure[]{\includegraphics[scale = 0.37]{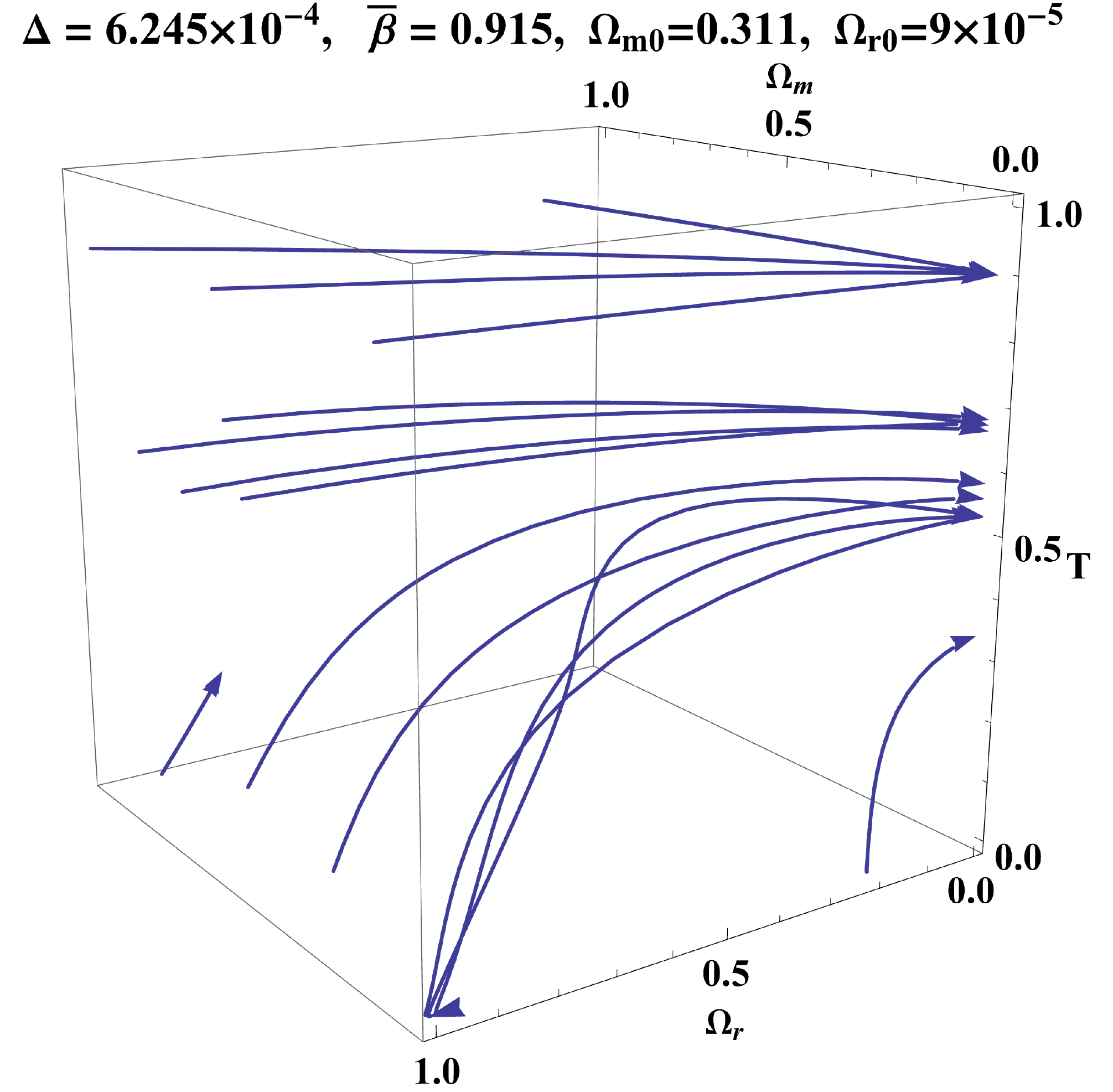}} 
     \subfigure[]{\includegraphics[scale = 0.37]{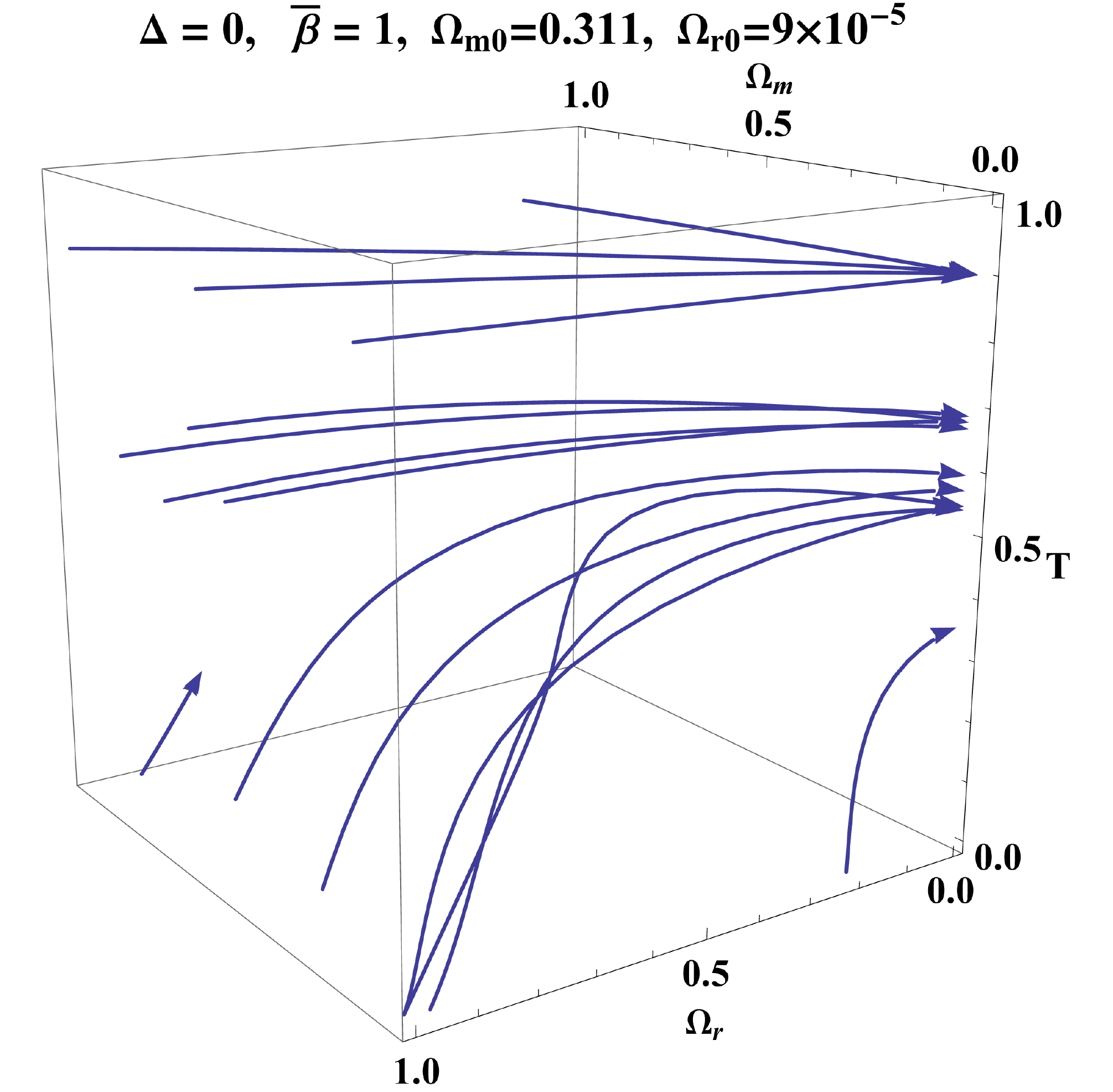}}
    \caption{\label{Fig7} Streamlines of the flow of \eqref{syst4.49} for (a) $\Delta=6.245\times 10^{-4}, \bar{\beta}= 0.915, \Omega_{m0}=0.311, \Omega_{r0}= 9\times 10^{-5}$ obtained from the joint constraints and (b) $\Delta=0, \bar{\beta}= 1, \Omega_{m0}=0.311, \Omega_{r0}= 9\times 10^{-5}$ for $\Lambda$CDM model.}
\end{figure*}
In Figure \ref{fig:my_label-3} are represented streamlines of the flow of  \eqref{eq:83} onto the unwrapped solution space (left panel) projected   both   onto the plane $\theta_1, \theta_2$ and  the torus $\mathbf{T}^1$, setting $T=1$ (right panel) of the  solution space of system  \eqref{eq:83} for the joint constraint value $\Delta=6.245\times 10^{-4}$, and the extreme value $\Delta=1$, respectively.

The results summarized in points 1-3 above are confirmed in  Figure \ref{fig:my_label-3}. 
That is, the lines of equilibrium points $dS_{++}$, $dS_{+-}$, $dS_{-+}$ and $dS_{--}$ (the family $dS$),  denoting  the de Sitter solutions, are stable.  The equilibrium points  $R_{+\pm}^{(0)}$, $R_{-\pm}^{(0)}$, $R_{+\pm}^{(1)}$ and $R_{-\pm}^{(1)}$ (the family $R$), associated to radiation dominated solutions for $\Delta=0$, are saddles. Finally, the equilibrium points 
$M_{\pm +}^{(0)}$, $M_{\pm -}^{(0)}$, $M_{\pm +}^{(1)}$ and $M_{\pm -}^{(1)}$ (the family $M$), associated to matter dominated solutions for $\Delta=0$,  are saddles.

In Figure \ref{Fig7} streamlines of the flow of \eqref{syst4.49} are presented. We select the values  (a) $\Delta=6.245\times 10^{-4}, \bar{\beta}=0.915, \Omega_{m0}=0.311, \Omega_{r0}= 9\times 10^{-5}$ obtained from the joint constraints, and (b) $\Delta=0, \bar{\beta}= 1, \Omega_{m0}=0.311, \Omega_{r0}= 9\times 10^{-5}$ for $\Lambda$CDM model. 

Figure \ref{Fig7}  shows, in a phase space, a crucial difference of  Barrow Entropy Cosmology (top panel) and $\Lambda$CDM  (bottom panel) related to the early universe. Barrow Entropy Cosmology does not admits a late-time radiation dominated phase (past attractor) and the solutions emerges from the point $(\Omega_r,\Omega_m, T)=(0,0,0)$ representing an effective DE- dominated early time attractor. However, both theories have the same late time dynamics, that is, the dominance of a de Sitter phase.  

Complementary, we define an unbounded variable  $\tilde{T}$ and keep $\theta_1, \theta_2$: 
\begin{equation}
 \tilde{T}=\frac{H_0}{H}, \quad     \theta_1=  \tan ^{-1}\left(\sqrt{\frac{8 \pi G \rho _{m}}{\Lambda
   }}\right),  \quad   \theta_2=  \tan ^{-1}\left( \sqrt{\frac{8 \pi G \rho _{r}}{\Lambda
   }}\right),
\end{equation}
and  using the logarithmic variable $\tau= \ln (a)=-\ln(1+z)$, we obtain the complementary system 
\begin{align}
\label{eq:5.44}
& \frac{d \tilde{T}}{d {\tau}}=\frac{\tilde{T} \left(3 \tan ^2(\theta_1)+4 \tan ^2(\theta_2)\right)}{(2-\Delta) \left(\tan ^2(\theta_1)+\tan ^2(\theta_2)+1\right)}, \nonumber\\
& \frac{d \theta_1}{d {\tau}}=-\frac{3}{4} \sin (2 \theta_1 ), \nonumber\\
& \frac{d \theta_2}{d {\tau}}=- \sin (2 \theta_2). 
\end{align}
Setting $a=1$ for the current universe, and considering the initial conditions: 
\begin{equation}
    \theta_1(0)= \tan ^{-1}\left(\sqrt{\frac{\Omega_{m0}}{\Omega_{\Lambda 0}}}\right), \quad \theta_2(0)= \tan ^{-1}\left(\sqrt{\frac{\Omega_{r0}}{\Omega_{\Lambda 0}}}\right), \quad \tilde{T}(0)=1,
\end{equation}
the system \eqref{eq:5.44} is integrated to obtain
\begin{align}
   & \theta_{1}(\tau )=\tan ^{-1}\left(e^{-3 \tau /2} \sqrt{\frac{\Omega_{m0}}{\Omega_{\Lambda 0}}}\right), \\
   & \theta_{2}(\tau )= \tan ^{-1}\left(e^{-2 \tau } \sqrt{\frac{\Omega_{r0}}{\Omega_{\Lambda 0}}}\right),\\
   & \tilde{T}(\tau )= e^{-\frac{4 \tau }{\Delta -2}} \left(\frac{\Omega_{\Lambda 0}+\Omega_{m0}+\Omega_{r0}}{\Omega_{\Lambda 0}}\right)^{\frac{1}{2-\Delta }}
   \left(\frac{e^{4 \tau } \Omega_{\Lambda 0}+e^{\tau } \Omega_{m0}+\Omega_{r0}}{\Omega_{\Lambda 0}}\right)^{\frac{1}{\Delta -2}}.
\end{align}
Finally, we  work out the exact evolution of $H$, $\rho_m$ and $\rho_r$:
\begin{align}
H & = H_0  \left(\frac{ \Omega_{\Lambda 0}+e^{-3\tau } \Omega_{m0}+e^{-4 \tau }\Omega_{r0}}{\Omega_{\Lambda 0}+\Omega_{m0}+\Omega_{r0}} \right)^{\frac{1}{2-\Delta}} \nonumber \\
& = H_0  \left(\frac{ \Omega_{\Lambda 0}+ \Omega_{m0}  (1+z)^3+ \Omega_{r0}  (1+z)^4}{\Omega_{\Lambda 0}+\Omega_{m0}+\Omega_{r0}} \right)^{\frac{1}{2-\Delta}}, \\
\rho_{m} & =\frac{\Lambda  \tan
   ^2(\theta_1)}{8 \pi  G}= \frac{\Lambda  e^{-3 \tau } \Omega_{m0}}{8 \pi  G \Omega_{\Lambda 0}}=\frac{3 H_0^2\Omega_{m0}}{8 \pi G} (z+1)^3,\\
 \rho_{r}   &=\frac{\Lambda  \tan
   ^2(\theta_2)}{8 \pi  G}= \frac{\Lambda  e^{-4 \tau } \Omega_{r0}}{8 \pi  G \Omega_{\Lambda 0}}=\frac{3 H_0^2\Omega_{r0}}{8 \pi G} (z+1)^4.
\end{align}
These expressions are used to calculate the fractional energy densities corresponding to matter,   radiation, and    a $\Lambda$-like source as follows: 
\begin{equation}
    \Omega_m(z)= \frac{8 \pi G \rho_m}{3 H^2}=  \Omega_{m0}  (z+1)^3 \left(\frac{ \Omega_{\Lambda 0}+ \Omega_{m0}  (1+z)^3+ \Omega_{r0}  (1+z)^4}{\Omega_{\Lambda 0}+\Omega_{m0}+\Omega_{r0}} \right)^{-\frac{2}{2-\Delta}},
    \label{eq:om_matrad}
\end{equation}
\begin{equation}
    \Omega_r(z)= \frac{8 \pi G \rho_m}{3 H^2} = \Omega_{r0}  (z+1)^4 \left(\frac{ \Omega_{\Lambda 0}+ \Omega_{m0}  (1+z)^3+ \Omega_{r0}  (1+z)^4}{\Omega_{\Lambda 0}+\Omega_{m0}+\Omega_{r0}} \right)^{-\frac{2}{2-\Delta}},
    \label{eq:or_matrad}
\end{equation}
and
\begin{equation}
    \Omega_\Lambda(z)= \frac{\Lambda}{3 H^2}=  \Omega_{\Lambda 0} \left(\frac{ \Omega_{\Lambda 0}+ \Omega_{m0}  (1+z)^3+ \Omega_{r0}  (1+z)^4}{\Omega_{\Lambda 0}+\Omega_{m0}+\Omega_{r0}} \right)^{-\frac{2}{2-\Delta}}, \quad  \Omega_{\Lambda 0}= \frac{\Lambda}{3 H_0^2}.
    \label{eq:ol_matrad}
\end{equation}
From  Eq. \eqref{ec:rhoDE} we  infer\
 \begin{align} 
\label{ec:omegaDE2}
&\Omega_{DE}(z)=\frac{8\pi G \rho_{DE}}{3 H^2}= \frac{\Lambda}{3H^2}+ \left[1-\frac{ \beta (\Delta+2)}{2-\Delta} 
H^{-\Delta}
\right]\nonumber \\
 & = 1+ \Omega_{\Lambda 0} \left(\frac{ \Omega_{\Lambda 0}+ \Omega_{m0}  (1+z)^3+ \Omega_{r0}  (1+z)^4}{\Omega_{\Lambda 0}+\Omega_{m0}+\Omega_{r0}} \right)^{-\frac{2}{2-\Delta}} \nonumber \\
 & - \frac{ (\Delta+2)}{(2-\Delta)} \frac{1}{\bar{\beta}}
\left(\frac{ \Omega_{\Lambda 0}+ \Omega_{m0}  (1+z)^3+ \Omega_{r0}  (1+z)^4}{\Omega_{\Lambda 0}+\Omega_{m0}+\Omega_{r0}} \right)^{-\frac{\Delta}{2-\Delta }}\nonumber \\
&= 1+ \Omega_{\Lambda 0} \left(\frac{ \Omega_{\Lambda 0}+ \Omega_{m0}  (1+z)^3+ \Omega_{r0}  (1+z)^4}{\Omega_{\Lambda 0}+\Omega_{m0}+\Omega_{r0}} \right)^{-\frac{2}{2-\Delta}} \nonumber \\
 & - \left(\Omega_{\Lambda0}+\Omega_{m0}+\Omega_{r0}\right)
\left(\frac{ \Omega_{\Lambda 0}+ \Omega_{m0}  (1+z)^3+ \Omega_{r0}  (1+z)^4}{\Omega_{\Lambda 0}+\Omega_{m0}+\Omega_{r0}} \right)^{-\frac{\Delta}{2-\Delta }}\nonumber\\
&= 1   - \left(  \Omega_{m0}  (1+z)^3+ \Omega_{r0}  (1+z)^4\right)
\left(\frac{ \Omega_{\Lambda 0}+ \Omega_{m0}  (1+z)^3+ \Omega_{r0}  (1+z)^4}{\Omega_{\Lambda 0}+\Omega_{m0}+\Omega_{r0}} \right)^{-\frac{2}{2-\Delta}},
 \end{align}
where we use Eq. 
\eqref{flatness2} to eliminate the term $\left(\frac{2+\Delta}{2-\Delta}\right)\frac{1}{\bar{\beta}}$,  recall $-\frac{\Delta}{2-\Delta}= 1 -\frac{2}{2-\Delta}$,  and  algebraically manipulate the equations to obtain $\Omega_{DE}=1- \Omega_m - \Omega_r$. It is straightforward to obtain Eqs. \eqref{eq:om_matrad}, \eqref{eq:or_matrad}, \eqref{eq:ol_matrad} and \eqref{ec:omegaDE2} in terms of the scale factor by replacing $z=(1/a) -1$.
\begin{figure}
\centering
\includegraphics[width=0.65\textwidth]{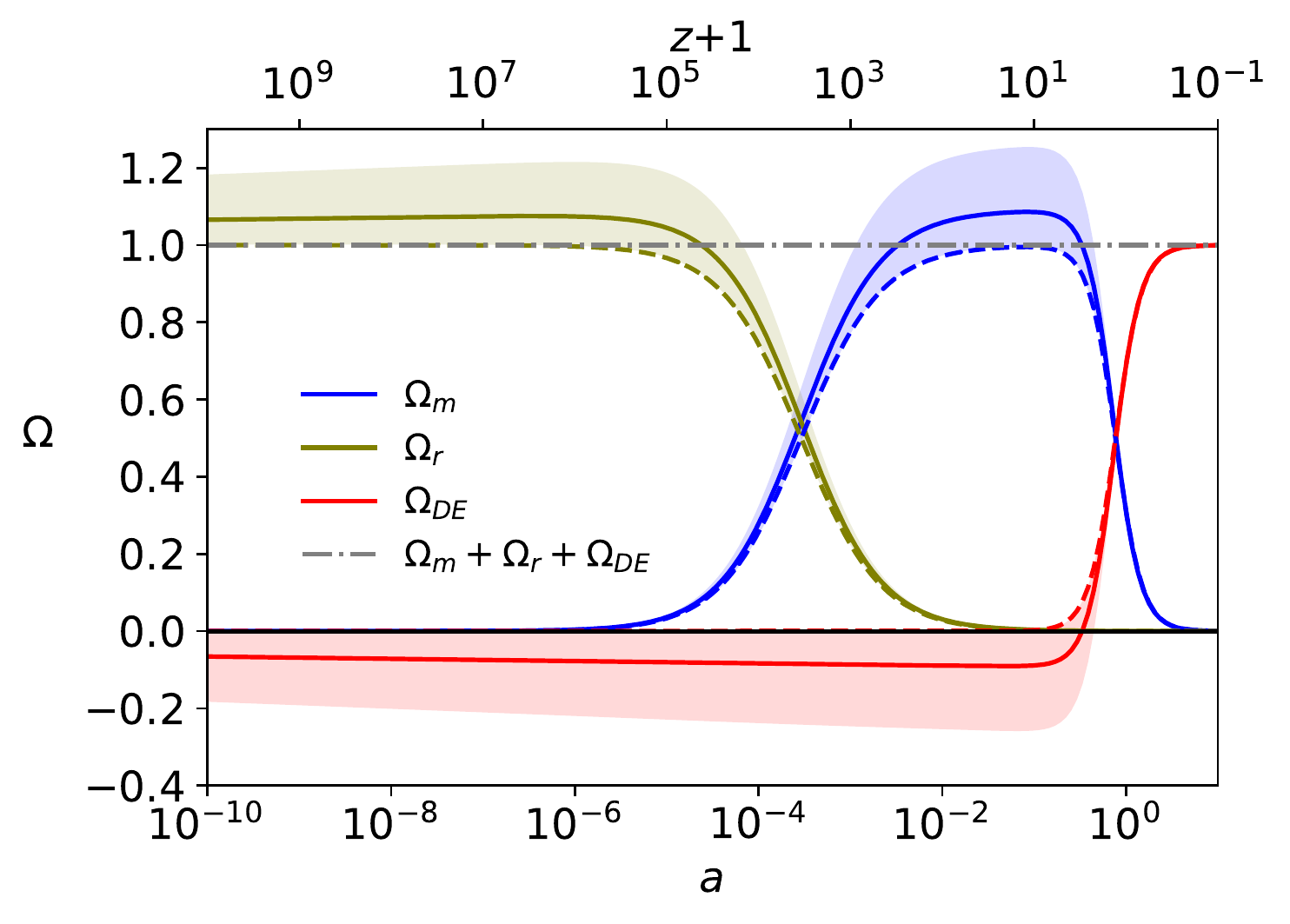}
\caption{Evolution of the density parameters for matter, radiation,  and effective dark energy using the joint constraint of Barrow Cosmology (solid lines) and the values for the standard model $\Delta=0$, and $\bar\beta=1$ (dashed lines). The shadow regions represent the $3\sigma$ confidence levels.
}
\label{fig:omegas_matrad}
\end{figure}
Figure \ref{fig:omegas_matrad} shows the evolution of the density parameters $\Omega_m$, $\Omega_r$  and $\Omega_{DE}$ vs the scale factor $a$ for two cases: the joint constraints for $h$, $\Omega_{m0}$, $\Delta$ and $\bar{\beta}$  shown in  Table \ref{tab:bestfits} (solid lines), and the values for the standard model $\Delta=0$, $\bar\beta=1$ (dashed lines).
We estimate $\Omega_{r0}\sim 9\times 10^{-5}$ consistent with the CMB data \cite{Aghanim:2018}. The shadowed regions represent the $3\sigma$ confidence levels. It is worthy to note that there are values of the energy densities which satisfy $\Omega_m>1$, $\Omega_r>1$, and $\Omega_{DE}<0$, however, these values are close to $\Lambda$CDM lines within the $3\sigma$ error propagation.

\section{Summary and discussion} \label{sec:Con}

Barrow entropy cosmology is a recent model \cite{Saridakis:2020lrg} based on the modification of the entropy-area black hole relation proposed by Barrow \cite{Barrow:2020tzx} that involves a new parameter $\Delta$,  recovering the standard form of the Bekenstein entropy for $\Delta=0$.
Considering this new relation, the modified Friedmann equations governing the dynamics of the Universe can be obtained from the gravity-thermodynamics approach. These new equations contain two parameters $\Delta$, and $\bar\beta$ (where  the standard model is recovered  for $\Delta=0$ and $\bar{\beta}=1$) and  could source the cosmic acceleration at late times. We investigate two Barrow cosmological models: I) Universe filled only by a matter component and II) Universe filled by matter and radiation components.
Furthermore, we divide the study of Barrow proposition in two ways: an observational approach and a dynamical system stability analysis.   

For the first approach we constrained the free parameters $\Delta$, $h$, and $\Omega_{m0}$, for both cosmological models,  employing  the observational Hubble data, type Ia supernovae, HII galaxies, strong lensing systems, baryon acoustic oscillations, and a joint analysis of these samples. We provide the observational constraints in section \ref{sec:data}  (see table \ref{tab:bestfits}), showing that for the model I the Barrow parameters are $\Delta=(5.912^{+3.353}_{-3.112})\times 10^{-4}$, $\bar{\beta}=0.920^{+0.042}_{-0.042}$, and for the model II $\Delta=(6.245^{+3.377}_{-3.164})\times 10^{-4}$, $\bar{\beta}=0.915^{+0.043}_{-0.043}$. For both models these  constraints are consistent at $2\sigma$ with  both the standard cosmological model and the standard entropy-area entropy relation. By reconstructing the cosmic expansion rate using the joint constraints in both cosmologies, we found consistency with the observational Hubble data. In addition, for the more realistic model with matter and radiation components, we   calculated the deceleration parameter and  obtained a transition at $z_{t}\simeq 0.711^{+0.035}_{-0.034}$ from a decelerated stage to an accelerated stage with $q_0=-0.573^{+0.019}_{-0.019}$, suggesting a de Sitter solution. We also confirm that, under these cosmologies, the equation of state of the effective dark energy can undergo from a quintessence-like regime to a phantom-like  one as found by \cite{Saridakis:2020lrg}, yielding $w_{DE}(0)\simeq -1.000134^{+0.000069}_{-0.000068}$ at current times which is consistent with the cosmological constant at $1\sigma$. 
Furthermore, we estimated the age of the Universe  as $t_{0}\simeq 14.062^{+0.179}_{-0.170}$ Gyr, consistent   within $2\sigma$ confidence level with the measurements of Planck  \cite{Aghanim:2018}.

The second approach, the stability analysis, allowed to find regions on the parameter space where the different cosmic epochs take place. In this regard, we obtained a qualitative description of the local and global dynamics of
both cosmological scenarios, irrespective of the initial conditions and the
specific evolution of the universe. Moreover, we have found asymptotic
solutions  calculated various theoretical values for the observable quantities that can be compared with previous observational constraints. From the analysis at the finite region of the phase space in the model I, we have found the line $A(\Omega_\Lambda)$ of equilibrium points,  which is is the attractor of the system  and represents the  de Sitter solutions. For $\Delta=0$, the line $B(\Omega_m)$ contains the past attractors, which represents the  matter dominated solutions.   Additionally, we have defined  the compact variables (assuming $H \geq 0, H_0>0$) based on the approach by  \cite{Alho:2014fha}. We have found two equivalent (due to the discrete symmetry) hyperbolic fixed points $M_{\pm}$ associated with dust fluid for $\Delta=0$, and two equivalent fixed points $dS_{\pm}$ corresponding  to the de Sitter states (see Table \ref{tab:my_label2}). 
On the other hand, in the model II, the line $A(\Omega_\Lambda)$ is the attractor of the system and represents the de Sitter solutions. For $\Delta=0$, the line $B(\Omega_m)$ is a saddle, indicating the matter dominated solutions,   while   the line $C(\Omega_r)$   contains  the past attractors and specify the radiation dominated solutions. Finally, we found  three  classes of equilibrium points/ lines:  $dS$,  $R$ and $M$; with $dS$ denoting  the de Sitter solutions, and  for $\Delta=0$,  $R$ and $M$  are  the radiation-dominated   and dust fluid  solutions, respectively  (see Table \ref{tab:my_label2B}). 

For both models we reconstruct the evolution of the density parameters $\Omega_m$, $\Omega_r$, and $\Omega_{DE}$ as a function of the scale factor $a$ for two cases: using the joint constraints for $h$, $\Omega_{m0}$, $\Delta$ and $\bar{\beta}$, and the values for the standard model $\Delta=0$, $\bar\beta=1$ (see Figs. \ref{fig:omegas_mat} and \ref{fig:omegas_matrad}). We found that at the early times there are values of the energy densities which satisfy $\Omega_m>1$, $\Omega_r>1$, and $\Omega_{DE}<0$. However, these values are close to $\Lambda$CDM lines within the $3\sigma$ error propagation. Thus, with more and high precision cosmological data, these non-physical density parameter values could be avoided. 

A crucial difference of  Barrow Entropy Cosmology  and the standard $\Lambda$CDM  model is related to the early universe. Barrow Entropy Cosmology does not admits a late-time radiation dominated phase  and the solutions are past asymptotic to a point representing an effective DE- dominated early time attractor. However, both theories have the same late time dynamics, that is, the dominance of a de Sitter phase.   In summary, we have showed, from several points of view, that the dynamical equations have a de Sitter solution at late times but the dynamics at early times is not consistent with the evolution of the standard cosmological model.

\section*{Acknowledgments} 

The authors are grateful for the figure \ref{Fig7} provided by Alfredo D. Millano (PhD student at Universidad Católica del Norte (UCN)). G.L. was funded by  Agencia Nacional de Investigaci\'on y Desarrollo - ANID for financial support through the program FONDECYT Iniciaci\'on grant no.
11180126 and by Vicerrectoría de Investigación y Desarrollo Tecnológico at UCN. J.M. acknowledges the support from ANID project Basal AFB-170002 and ANID REDES 190147. M.A.G.-A. acknowledges support from SNI-M\'exico, CONACyT research fellow, ANID REDES (190147), COZCyT and Instituto Avanzado de Cosmolog\'ia (IAC). A.H.A. thanks to the PRODEP project, Mexico for resources and financial support and thanks also to the support from Luis Aguilar, Alejandro de Le\'on, Carlos Flores, and Jair Garc\'ia of the Laboratorio Nacional de Visualizaci\'on Cient\'ifica Avanzada. V.M. acknowledges support from Centro de Astrof\'{\i}sica de Valpara\'{i}so and ANID REDES 190147.


\providecommand{\href}[2]{#2}\begingroup\raggedright\endgroup


\end{document}